# Hole-doping reduces the coercive field in ferroelectric hafnia

Pravan Omprakash[1,*], Gwan Yeong Jung[2], Guodong Ren[1], Rohan Mishra[2,1,†]

[1]*Department of Mechanical Engineering & Material Science, Washington University in St. Louis, St. Louis, MO 63130, USA*

[2]*Institute of Materials Science & Engineering, Washington University in St. Louis, St. Louis, MO 63130, USA*

[*]Corresponding author: o.pravan@wustl.edu

[†]Corresponding author: rmishra@wustl.edu

**ABSTRACT**

Ferroelectric hafnia ($HfO_2$) holds promise for next-generation memory and logic applications because of its CMOS compatibility. However, the high coercive field required for polarization switching in $HfO_2$ remains a critical challenge for efficient device operations. Using first-principles calculations and phenomenological modeling, we predict that hole doping can reduce the coercive field from 8 MV/cm in undoped hafnia to 6 MV/cm in hafnia doped with 0.2 holes per formula unit (f.u.). In the absence of doping, the reversal of polarization of the $Pca2_1$ phase is preferred through the non-polar, tetragonal $P4_2/nmc$ phase. This switching pathway involves the coupling of three hard distortion modes that render undoped hafnia as an improper ferroelectric. The overall energy barrier through this pathway remains unchanged (~80 meV/f.u.) upon hole doping. However, the introduction of holes hardens the polar $\Gamma_2^-$ distortion mode that connects the polar $Pca2_1$ phase to the non-polar, orthorhombic *Pbcm* phase, and reduces the energy barrier from 180 meV/f.u. in undoped hafnia to 80 meV/f.u. at 0.2 holes/f.u.. The activation of the latter switching pathway through the *Pbcm* phase can lead to a reversal in the polarization direction.

Overall, hole doping makes the switching pathway through the *Pbcm* phase competitive, and renders hafnia as a proper ferroelectric with a lower coercive field.

# I. Introduction

Hafnia ($HfO_2$) is used as the gate-dielectric layer in CMOS devices. The discovery of ferroelectricity in ultrathin films of hafnia has opened up prospects for the seamless integration of memory and logic [1,2]. However, the electric-field switching of polarization in ferroelectric hafnia is associated with large coercive fields ($E_C$ > 1 MV/cm) [3-5] due to its intrinsically high switching barriers, which can be a hindrance to energy-efficient device operations. Considering $HfO_2$ exists in multiple polymorphs, there is scope for reducing $E_C$ by activating alternate switching pathways that pass through one of its many metastable polymorphs. One such proposal, based on theoretical calculations, involves the use of epitaxial tensile strain, which is predicted to lower the energy of a centrosymmetric metastable phase, and decrease the switching barrier — measured as the depth of the theoretically calculated ferroelectric double well potential — from 0.8 to 0.55 eV [6,7]. However, considering that the vast majority of hafnia thin films are polycrystalline in nature, except for one recent report on the growth of single crystalline hafnia thin films [8], precise control over strain in experiments seems challenging. As an alternative, hole doping has been theoretically predicted [9,10] and experimentally demonstrated [11-14] to stabilize the metastable, polar orthorhombic phase ($Pca2_1$) of hafnia over its nonpolar monoclinic ($P2_1/c$) ground state. There are multiple ways to introduce holes into a ferroelectric, including through dopant elements [12,13,15], oxygen vacancies [16-18], electrostatic gating [19], and charged domain walls [20,21]. However, the impact of hole doping on the atomic-scale switching mechanisms in hafnia, and $E_C$, remains poorly understood.

There are two competing pathways to switch the direction of polarization in the $Pca2_1$ phase: the 'Shift Inside' (SI) pathway that takes the system through an intermediate non-polar $P4_2/nmc$ phase [7,22], and the 'Shift Across' (SA) pathway through an intermediate non-polar

*Pbcm* phase [7,22,23]. In pristine hafnia, without strain or doping, the SI pathway has been theoretically proposed as the preferred pathway with an energy barrier of 80 meV/f.u.. Apart from these two pathways, a handful of recent studies have proposed the transition between the antiferroelectric *Pbca* phase and the polar phase [24,25], which we refer to as the AFE-FE transition, a third pathway to switch the polarization. The antiferroelectric *Pbca* phase has a lower energy than the polar $Pca2_1$ phase, and can be stabilized with epitaxial strain [25]. Ferroelectricity is activated by overcoming the energy barrier between the two phases [24]. Previous studies have shown that dopants [23,26] or oxygen vacancies [10,18,27,28] can modify the $E_C$ for the SI pathway, while lesser attention has been paid to the effect of doping on the SA and AFE-FE pathways. In particular, the effect of intrinsic hole doping remains unexplored. An understanding of the role of hole doping could not only provide control over the $E_C$, it could also help explain switching characteristics in the presence of charged domain walls and defects [29].

In this Article, we show that hole doping can result in a transition in the switching route by stabilizing the SA pathway, thereby decreasing the $E_C$ by up to 14 % (6 MV/cm) at 0.2 holes per formula unit. We used a combination of density functional theory (DFT) calculations, group theoretical methods, and phenomenological modelling to study the modulation of switching pathways under hole doping. We find that hole doping reduces the energy barrier of the SA pathway by activating a polar $\Gamma_2^-$ mode and stabilizing the *Pbcm* phase. The activation of the SA pathway also leads to a reversal in the polarization direction, which changes the direction of domain wall movement [4] and the sign of the piezoelectric coefficient [7]. Conversely, the SI pathway remains largely unaffected by hole doping, which we attribute to its improper ferroelectric nature, as quantified by the coupling between hard phonon modes. The AFE-FE transition also shows a minimal increase in energy barrier (~10%) due to its improper nature. The AFE-FE transition also shows a minimal increase in energy barrier (~10%) due to its improper nature. The AFE-FE transition has been tied to the wake-up effect observed effect in $HfO_2$ samples, wherein a critical electric field of up to 0.8 MV/cm is required to activate ferroelectric response in the samples [24]. The slight increase in the AFE-FE energy barrier can thus explain the reason for hole-doped $HfO_2$ samples to also exhibit a pronounced wake-up effect [30]. Overall, we propose that hole doping could be a viable pathway to lower the switching voltage in hafnia for efficient device operations.

## II. Computational Details

We performed DFT calculations using the Vienna Ab-initio Simulation Package (VASP) [31]. We used projector augmented-wave (PAW) potentials and the Perdew-Burke-Ernzerhof (PBE) functional [32,33]. We used an energy cutoff of 600 eV. The Brillouin zone was sampled using a $4 \times 4 \times 4$ $k$-points grid for ionic and volume optimization and an $8 \times 8 \times 8$ grid for electronic structure calculations. The convergence criteria were set to $10^{-6}$ eV for energy and $10^{-3}$ eV/Å for forces, respectively. We use a gaussian smearing with a width of 0.008 eV. Electronic structure analysis was performed using the pymatgen package [34] and crystal structure visualization was done using VESTA [35].

The background charge method in VASP was used to simulate the effect of adding charge carriers in the system without introducing dopant atoms. In this approach, to ensure the charge neutrality in the presence of free carriers, a uniform background with opposite charge is added [36]. Symmetry-mode analyses were performed using the Bilbao crystallographic server [37] and the ISODISTORT package [38].

It has been theoretically proposed that the spontaneous polarization in hafnia depends on the switching pathway through either of the two non-polar reference phases, $P4_2/nmc$ and *Pbcm* [4,7]. Accordingly, we calculated the spontaneous polarization for both the pathways in undoped hafnia using the Born effective charges — that were themselves calculated using density functional perturbation theory (DFPT) — and multiplying them with the polar displacements in the $Pca2_1$ phase with respect to the two non-polar references phases. Hole-doping introduces states at the Fermi energy. Since polarization is ill-defined for metallic systems, we determined the polarization at different hole-doping concentrations, following the approach used by Li et al. [39]:

$$\overrightarrow{P_{\mathrm{s}}} = \frac{\mathrm{e}}{\mathrm{V}} \sum_{\mathrm{i}} (Z_i^* - \sigma) \vec{d}, \qquad (1)$$

where $Z_i^*$ is the Born effective charge for an oxygen ion, $i$, that undergoes off-centering displacements, $\vec{d}$, with respect to the chosen non-polar reference phase, V is optimized volume of the polar phase and $\sigma$ is localized charge on the displaced O ions. The localized charge upon introducing holes was calculated by integrating the unoccupied valence band density of states projected onto the displaced ions. This is used to correct the Born effective charge as shown in Eq. (1) to obtain the polarization. Using these polarization values, we computed $E_{\mathrm{c}}$, following Reyes-Lillo [40] et al.:

$$E_{\mathrm{c}} = \frac{E_{\mathrm{b}}}{V \cdot \overrightarrow{P_{\mathrm{s}}}}, \qquad (2)$$

where $E_b$ is the energy barrier for the selected pathway and $V$ is the volume of the optimized structure at the barrier. To benchmark our approach, we computed the energy barriers and $E_{\mathrm{c}}$ for undoped $ZrO_2$ following Reyes-Lillo et al. [40], and obtained a similar value of 4 MV/cm (Fig. S3 in the Supplemental Material [41]).

## III. Results

### A. Polarization switching pathways in $HfO_2$

As discussed in the Introduction, there are two intrinsic pathways to reverse the direction of polarization of the $Pca2_1$ phase, either through the $P4_2/nmc$ phase or the *Pbcm* phase, both of which are non-polar phases. We begin by examining the structure of the polar orthorhombic $Pca2_1$ phase, which we refer to as the *o*-1 phase. The *o*-1 phase has two different oxygen sub-lattices, one having three-fold coordination, and the other having four-fold coordination. The *o*-1 phase can adopt two different polarization configurations, as shown in Fig. S1 in the Supplemental Material [41]. These configurations differ by the position of the three-fold coordinated oxygen atoms ($O_I$, marked in blue in Figure 1)). The four-fold coordinated $O_{II}$ atoms (marked in red) are located in a centrosymmetric position with respect to the four neighboring Hf atoms, and do not undergo displacements during the switching. In contrast, the $O_I$ sub-lattice is off-centered with respect to the Hf sub-lattice, and results in the polarization. In the SI pathway, the $O_I$ atoms shift within the unit cell, i.e., they do not cross the Hf-planes to attain a centrosymmetric position. The SI pathway is indicated by green arrows in Fig. S1 in the Supplemental Material [41]. Conversely, in the SA pathway, the $O_I$ atoms move towards opposite directions to align with the Hf planes on either side, and result in the centrosymmetric *Pbcm* phase (*o*-4 phase). The $O_I$ atoms subsequently move across the Hf planes to adopt the switched configuration with its polarization reversed, as shown by the orange arrows in Fig. S1 in the Supplemental Material [41]. The opposite displacements of the $O_I$ atoms and the resulting non-polar reference phases mark the difference between the SA and the SI pathways [4,7].

The above displacements are caused by different symmetry-lowering distortions from the non-polar reference phases and render hafnia as either a proper or improper ferroelectric [42,43]. Proper ferroelectrics have a zone-center polar instability that appears as a *soft* phonon mode with imaginary frequencies. In the SA pathway, the non-polar *o*-4 phase is dynamically unstable, exhibiting three soft phonon modes; the zone-center $\Gamma_2^-$ mode that leads to the polar *o*-1 phase, as shown in Fig. 1(a), the $\Gamma_4^+$ mode leading to the non-polar monoclinic phase, and the zone-boundary $X_2^-$ mode to the antipolar *Pbca* phase, as shown with orange arrows in Fig. S4 in the Supplemental Material [41]. Thus, $HfO_2$ when switching via the SA pathway can be considered as a proper ferroelectric, as has been noted by Aramberri et al. [43]. The spontaneous polarization was calculated to be 75 μC/cm$^2$, which is consistent with a previous report [4].

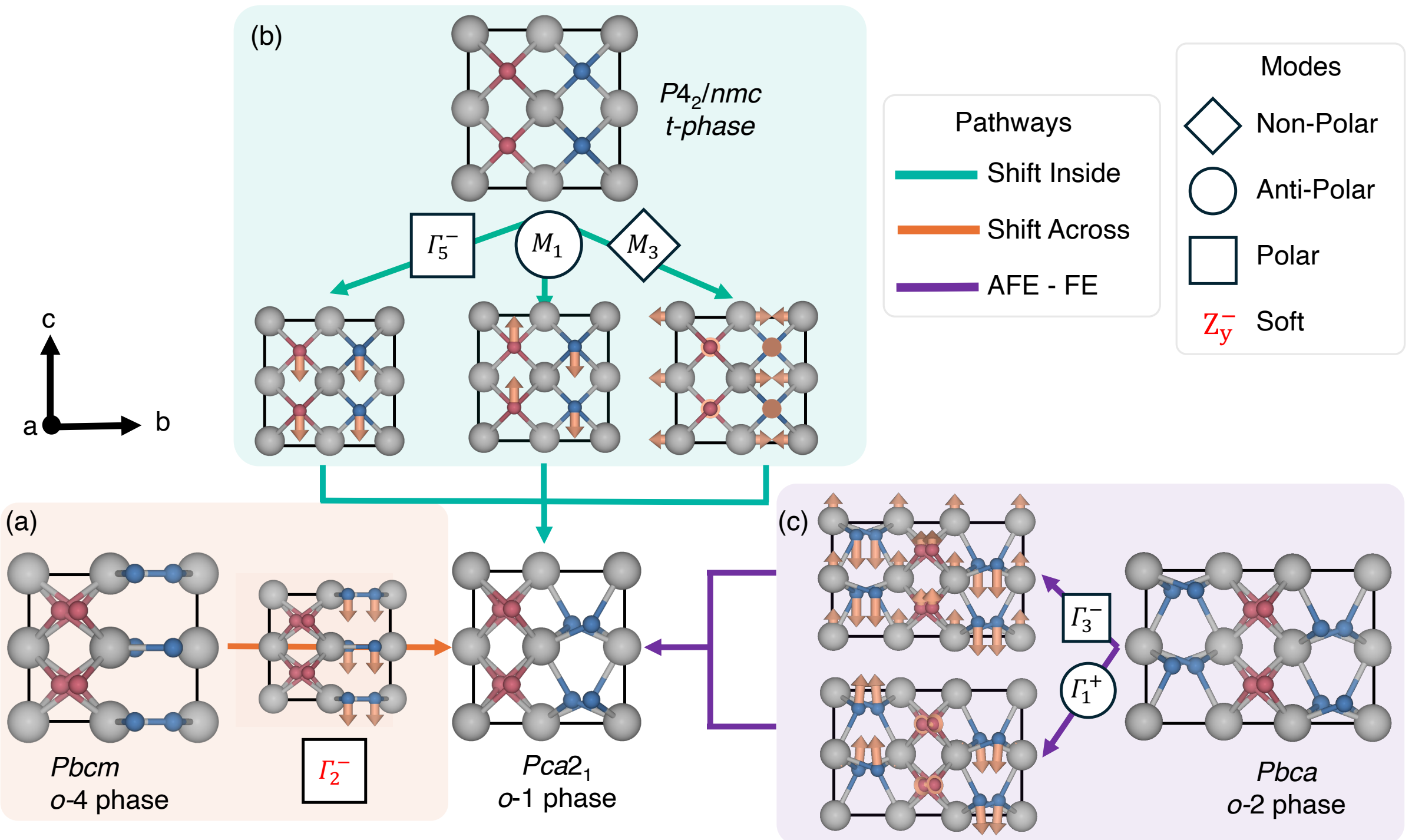

**Fig 1. Schematics of the three polarization switching pathways from the polar orthorhombic *Pca*2₁ phase along with the associated distortion modes.** (a) The shift-across (SA) pathway, in orange, originates from the *Pbcm* (*o*-4 phase) phase, and is driven by the zone-center, soft-polar $\Gamma_2^-$ mode. (b) The shift-inside (SI) pathway, in green, originates from the $P4_2/nmc$ (*t*-phase) phase, and consists of three coupled hard modes: polar $\Gamma_5^-$, antipolar $M_1$, and a non-polar $M_3$ mode. (c) The AFE-FE transition, in purple, originates from the antiferroelectric

*Pbca* (*o*-2 phase) phase, and is similar to the SI pathway with two hard modes (polar $\Gamma_3^-$ , antipolar $\Gamma_1^+$ modes) coupled together.

In the SI pathway, the tetragonal $P4_2/nmc$ phase (*t*-phase) possesses no soft modes and is dynamically stable [44]. There are three hard modes that couple together to lead to the *o*-1 phase [42]. The $\Gamma_5^-$ mode is a polar distortion that corresponds to parallel movement of the $O_I$ and $O_{II}$ atoms along the *c*-axis. The $M_I$ mode is an antipolar distortion that consists of antiparallel movement of the $O_I$ and $O_{II}$ atoms along the *c*-axis. Additionally, the $M_3$ mode is non-polar and involves $O_I$ and $O_{II}$ displacements along the *a*-axis. These three modes are summarized in Fig. 1(b). The spontaneous polarization for this pathway was determined to be 55 μC/cm$^2$, which is consistent with previous reports [4,44]. The coupling of these three hard modes in the SI pathway can render hafnia as a hybrid improper ferroelectric [42].

Proper and improper ferroelectric systems can have different electronic and thermal properties. Since improper ferroelectrics are not driven solely by a polar mode, they can couple with other phenomena such as strain and magnetism, influencing their ferroelastic and multiferroic properties [45-47]. Improper ferroelectrics usually have higher coercive fields because of the coupling between multiple modes, while proper ferroelectrics show higher permittivity [43]. Furthermore, improper ferroelectrics show sustained polarization even in ultrathin films [48,49], while proper ferroelectrics are affected by depolarizing fields, and, below a critical thickness, their polarization is quenched [50]. Thus, the ability to select either the SA or SI pathway in ferroelectric hafnia can enable control over its electronic, thermal and elastic properties.

Recent experiments have reported a transition from the AFE *Pbca* phase (*o*-2 phase) to the polar *o*-1 phase [24,25,51]. The *o*-2 phase can be stabilized over the ground-state *m*-phase with epitaxial strain and/or chemical doping [25], and has a lower energy of 10 meV/fu than the *o*-1 phase. To activate ferroelectricity in $HfO_2$, the *o*-1 phase has to then be accessed, either with an electric field or stresses strong enough to surmount the energy barrier associated with the AFE-FE transition. This process is directly tied to the wake-up effect in $HfO_2$ [24], wherein a minimum critical electric field is required to obtain a ferroelectric response. The *o*-2 phase, being antipolar, has double the periodicity to the *o*-1 phase, as shown in Fig. 1(c). The AFE-FE transition can occur through two pathways, differentiated in a similar manner as the SA and SI pathways. The barriers for both the pathways at 0 K and higher temperatures have been discussed by Tan et al. [52]. The pathway with oxygen movements across the Hf plane (called the T-pathway) was found to have a lower energy barrier than the pathway with oxygen movements within the unit cell (called the N-pathway). Here, we focus on the lower energy T-pathway, which can be described by the symmetry modes connecting the *o*-2 phase and *o*-1 phase. The *o*-2 phase has two coupled hard modes – the polar $\Gamma_3^-$ mode corresponding to the displacement of the $O_I$ sub-lattice along the *c*-axis, and the antipolar $\Gamma_1^+$ mode resulting in antiparallel displacements of the $O_I$ sub-lattice along the *c*-axis. The AFE-FE transition is analogous to the SI pathway; both driven by the coupling of hard modes. Consequently, we find the AFE-FE transition and the SI pathway respond similarly to hole doping, while the soft mode in the SA pathway couples differently with the added holes.

### B. Effect of hole doping on the SA pathway

To determine the response of the SA pathway to hole doping, we calculated the change in switching barrier with increasing hole concentration. We used a Landau model described using Eq. (3), to capture the effect of hole doping on the characteristics of the phase transition:

$$E = a_{200} Q_P^2 + a_{400} Q_P^4 + a_{600} Q_P^6. \quad (3)$$

The SA pathway consists of a soft polar mode $\Gamma_2^-$ and a symmetry-preserving mode $\Gamma_1^+$, which are considered together with an amplitude denoted by $Q_P$. For six different doping concentrations of [0, 0.04, 0.08, 0.12, 0.16, and 0.20] holes per formula unit (h/f.u.), we generated 10 frozen intermediate images between the non-polar *o*-4 phase and the polar *o*-1 phase along the SA pathway by freezing the $\Gamma_2^-$ and $\Gamma_1^+$ modes together and performed static DFT calculations to obtain the energy. In Eq. (3), the coefficients $a_{200}$, $a_{400}$, and $a_{600}$ are obtained by fitting to DFT energies using standard polynomial fitting algorithms as implemented in NumPy. For up to 0.12 h/f.u., the SA pathway exhibits a double-well potential similar to that observed in proper ferroelectrics, as shown in Fig. 2(a). In the undoped case, the energy barrier is 180 meV/f.u., owing to the highly unstable *o*-4 phase. Here, the energy barrier is defined as the energy difference between the *o*-1 phase, at the bottom of the well, and the maximum value along the path towards the *o*-4 phase. Introduction of holes lowers the switching barrier, as shown in the upper panel of Fig. 2(b). The decrease corresponds to the hardening of the soft polar mode ($\Gamma_2^-$). As the soft mode hardens, the energy gained by freezing the mode decreases, which is shown in the lower panel of Fig. 2(b). This in turn, leads to the *o*-4 phase becoming more stable with hole doping, as shown in Fig. S5 in the Supplemental Material [41].

What stabilizes at doping concentration and effect of volume constraint, etc. then conclusion from result.

Beyond 0.12 h/f.u., the energy profile of the SA pathway develops a hull, forming a triple-well potential, which is described in the Landau model in Eq. (3) by an increase in the magnitude of the higher order coefficients as well as a change in sign—from positive to negative—of the $a_{400}$ coefficient. This change in sign, indicates a change from instability to metastability of the *o*-4 phase. The magnitude and sign of each coefficient as a function of hole doping are plotted in Fig.

S7 in the Supplemental Material [41]. Nevertheless, the overall energy barrier for the pathway continues to decrease, reaching ~80 meV/f.u. at 0.2 h/f.u. Thus, the suppression of the soft mode in the $o$-4 phase leads to its stabilization and reduces the energy barrier.

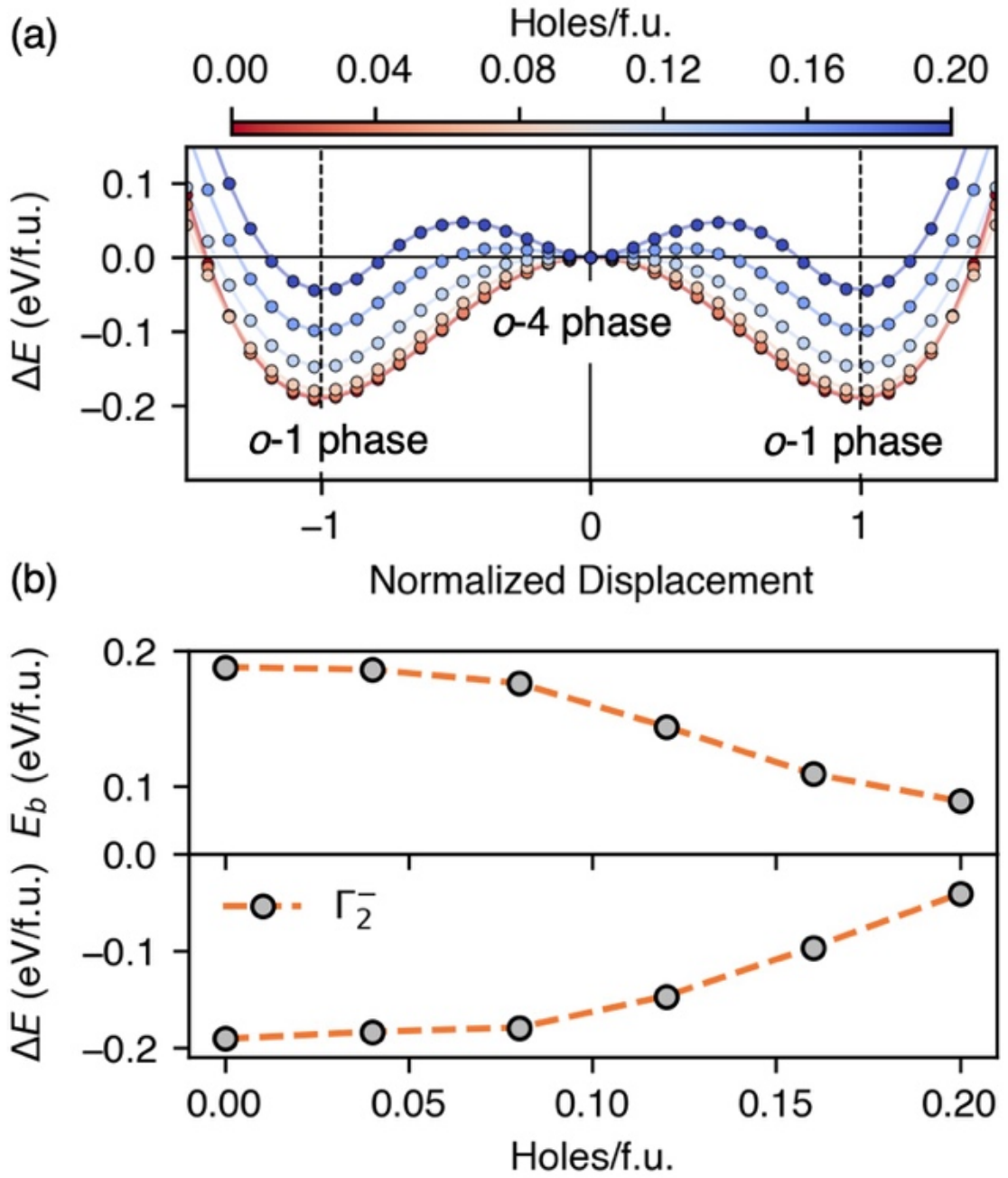

**Fig 2. Modulation of the Shift Across (SA) pathway with hole doping.** (a) The change in energy along the SA pathway for polarization reversal with increasing hole doping. (b) The corresponding change in energy barrier is shown in the top panel, with the change in energy between the $o$-4 phase and $o$-1 phase shown in the bottom panel, which can also be seen as a hardening of the soft polar mode, $\Gamma_2^-$.

### C. Effect of hole doping on the SI pathway

Unlike the SA pathway, the SI pathway involves the coupling of the polar $\Gamma_5^-$ mode with antipolar $M_1$ and non-polar $M_3$ modes, as shown in Fig. 1(b). In undoped $HfO_2$, the SI pathway shows a lower energy barrier of 80 meV/f.u, compared to the SA pathway. Hole-doping has a negligible impact, causing only a small increase of ~5 meV/f.u. in the energy barrier at higher doping levels of (0.16 – 0.2) h/f.u., as shown in Fig. 3(a). By contrast, the barrier for the SA pathway decreases by ~100 meV/f.u. under the same range of doping levels. Thus, the SI pathway is less sensitive to hole doping when compared to the SA pathway. To understand this minimal change, we decomposed the free-energy expression for the SI pathway into the three phonon modes and the energy contributions from their couplings, as described in Eq. (4):

$$\begin{aligned} E = {} & a_{200}Q^2_{\Gamma_5^-} + a_{020}Q^2_{M_1} + a_{002}Q^2_{M_3} + a_{400}Q^4_{\Gamma_5^-} + a_{040}Q^4_{M_1} + a_{004}Q^4_{M_3} + a_{600}Q^6_{\Gamma_5^-} + \\ & a_{060}Q^6_{M_1} + a_{006}Q^6_{M_3} + a_{220}Q^2_{\Gamma_5^-}.Q^2_{M_1} + a_{022}Q^2_{M_1}.Q^2_{M_3} + a_{202}Q^2_{\Gamma_5^-}.Q^2_{M_3} + \\ & a_{240}Q^2_{\Gamma_5^-}.Q^4_{M_1} + a_{024}Q^2_{M_1}.Q^4_{M_3} + a_{204}Q^2_{\Gamma_5^-}.Q^4_{M_3} + a_{420}Q^4_{\Gamma_5^-}.Q^2_{M_1} + a_{042}Q^4_{M_1}.Q^2_{M_3} + \\ & a_{402}Q^4_{\Gamma_5^-}.Q^2_{M_3} + a_{111}Q_{\Gamma_5^-}.Q_{M_1}.Q_{M_3} + a_{222}Q^2_{\Gamma_5^-}.Q^2_{M_1}.Q^2_{M_3} + a_{113}Q_{\Gamma_5^-}.Q_{M_1}.Q^3_{M_3} + \\ & a_{131}Q_{\Gamma_5^-}.Q^3_{M_1}.Q_{M_3} + a_{311}Q^3_{\Gamma_5^-}.Q_{M_1}.Q_{M_3}. \end{aligned} \tag{4}$$

For the SI pathway, the three irreps ($Q_{\Gamma_5^-}, Q_{M_1}, Q_{M_3}$) were fitted using Eq. (4), following Delodovici et al. [42]. The resulting coefficients are tabulated in Table S1.

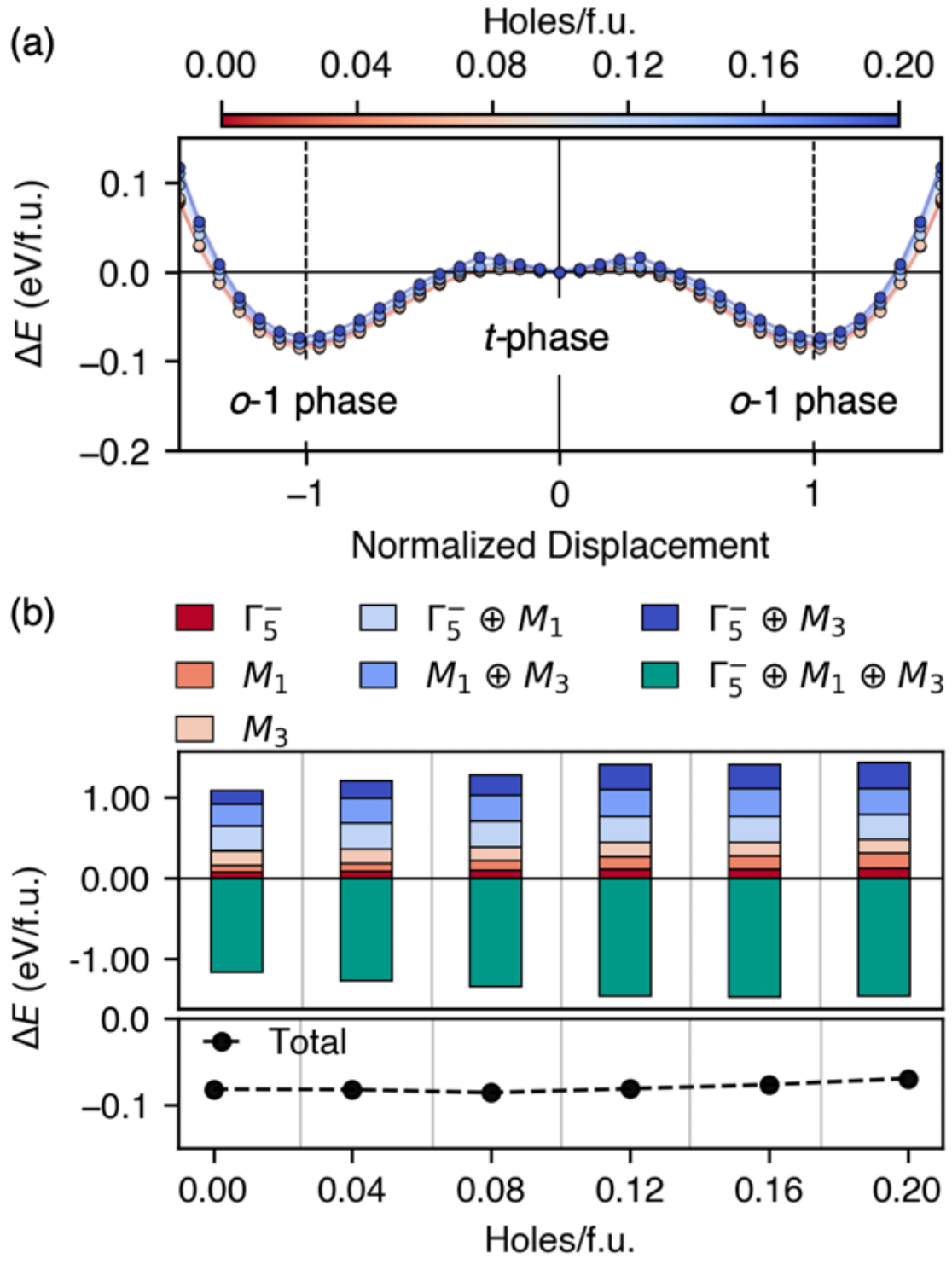

**Fig 3. The Shift Inside (SI) pathway is insensitive to hole doping.** (a) Energy landscape along the SI pathway remains unchanged with increasing hole doping. (b) The energy contributions of the three individual distortion modes, $\Gamma_5^-, M_1, and\ M_3$, the three doublet coupling terms, and the one triplet coupling term, as a function of hole doping are shown in the upper panel. The lower panel shows the energy barrier obtained by adding all the contributions for the different doping levels.

The energy contributions from the modes in Eq. (4) can be categorized into three parts—three single mode terms, three binomial coupling terms, and a trinomial coupling term. These individual contributions as a function of hole doping are plotted in the upper panel of Fig. 3(b). The single-mode terms and the binomial coupling terms contribute positively to the total energy of the *t*-phase, and thus do not stabilize the *o*-1 phase. In contrast, the trinomial coupling is the dominant energy contribution that results in the stabilization of the *o*-1 phase [42]. With increasing hole doping, the positive energy contributions increase but they are nearly counterbalanced by an increasingly negative trinomial contribution, leading to a negligible net change in the energy difference between the *t*-phase and the *o*-1 phase. The free energy pathways for each of these contributions are plotted in Fig. S8 in the Supplemental Material [41]. The hybrid improper ferroelectric nature of the SI pathway thus allows for preserving ferroelectricity in $HfO_2$ even when the holes are introduced at higher concentrations of up to 0.2 h/f.u.

**D. Effect of hole doping on the AFE-FE pathway**

We next take a look at the energy barriers for the AFE-FE pathway at different doping concentrations. As described in Section A, the AFE-FE transition includes a polar $\Gamma_3^-$ mode and the antipolar $\Gamma_1^+$ mode, both of which are hard modes, and couple to stabilize the polar *o*-1 phase. We calculated the energy barriers for the AFE-FE pathway at six doping concentrations, as shown in the left panel in Fig. 4(a). The high barrier of 180 meV/f.u. in the undoped system, which is consistent with a previous report [24], increases slightly to 198 meV/f.u. (~10%) with hole doping. In single crystalline $La_{0.1}Hf_{0.9}O_2$ thin films, epitaxial strain has been shown to stabilize the *o*-2 phase in the ultrathin limit of 10 nm and below, rather than stabilizing the *o*-1 phase or other non-polar phases [25]. Consequently, to activate ferroelectricity, accessing the *o*-1 phase requires overcoming the high energy barrier of >180 meV/f.u. between the *o*-2 phase and the *o*-1 phase, which has been termed as a wake-up effect. Such wake-up effects have also been observed in doped $HfO_2$-based ferroelectric films, including *p*-type dopants such as Gd, Y, and La, and aliovalent dopants Sr, and Zr [30,53,54]. The persistence of the high energy barrier of the AFE-FE transition in response to hole doping, as shown in Fig. 4(a), can explain the experimental observations of the wake-up effect in the doped films.

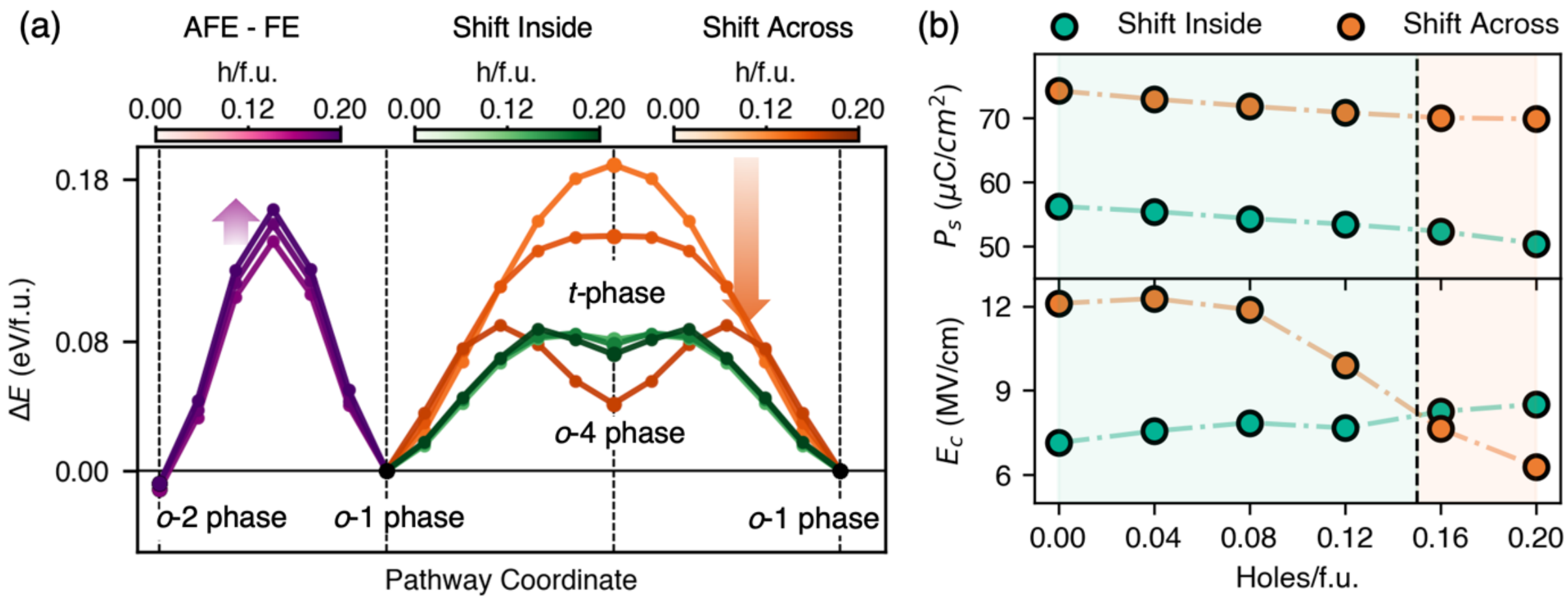

**Fig 4. The activation of the Shift Across pathway and lowering of $E_c$ with hole doping above 0.15 holes/f.u.** (a) The change in energy barrier for the AFE-FE transition along the *o*-2 phase to *o*-1 phase pathway (purple) with increasing hole concentration. A comparison of the change in the energy barriers between *o*-1 phases with opposite polarization along the SI (green) and SA (orange) pathways, through the *t*-phase and the *o*-4 phase, respectively. (b) The change in spontaneous polarization for the SI and SA pathways (top panel) along with the corresponding change in the coercive fields (bottom panel) with hole doping.

## E. Reduction in the coercive field by activating the SA pathway

Since the three pathways behave differently in the presence of holes, we have compared all the energy barriers together in Fig. 4(a). Starting with the lower energy $o$-2 phase (AFE), as has been observed in as-grown single crystalline thin films of La-$HfO_2$ [25], an AFE-FE transition is required to access the $o$-1 phase (FE). Once the FE phase is accessed, the polarization can be switched through either the SI or the SA pathway. The energy barrier of the SA pathway reduces from 180 meV/f.u. to 80 meV/f.u. at 0.2 h/f.u. and becomes competitive with the SI pathway. The energy barriers obtained from the symmetry-mode analysis described in Sections B and C are compared with solid-state nudged elastic band calculations for select doping levels of 0.0 and 0.2 h/f.u. for both the SI and SA pathways. The change in the pathways is shown in Fig. S9. The energy barrier does not change for the SI pathway and remains around 80 meV/f.u. Meanwhile, for the SA pathway, the barrier reduces to 40 meV/f.u. at 0.2 h/f.u. Therefore, the same trends exist between the symmetry-mode analysis and NEB calculations—hole doping reduces the energy barrier of the SA pathway and activates the pathway at higher doping levels.

While, the energy barrier for switching cannot be measured directly, the $E_C$, which is proportional to the energy barrier over the spontaneous polarization [7,55], can be obtained from the experimental hysteresis loop of a ferroelectric sample.

To calculate $E_c$, the spontaneous polarization ($\boldsymbol{P_s}$) must be first determined (for details on calculating $\boldsymbol{P_s}$, refer to the Methods section). We observe that the polar distortions in the system reduce by < 0.02 Å (refer to Fig. S9 in the Supplemental Material [41]) for the range of hole concentrations used here, indicating that $\boldsymbol{P_s}$ in $HfO_2$ is robust to hole doping, in agreement with previous theoretical calculations [18,56]. The change in the magnitude of $\boldsymbol{P_s}$ with increasing hole doping, calculated using Eq. (1), is plotted for the SA and SI pathways in the upper panel of Fig. 4(b). The polarization in both the pathways does not decrease significantly (< 10 μC/cm$^2$) and the $\boldsymbol{P_s}$ for the SA pathway is higher than the SI pathway by ~15 μC/cm$^2$.

Using the polarization from Fig. 4(c), the $E_\mathrm{c}$ for the SI and SA pathways were calculated using Eq. (2) and are plotted in the lower panel of Fig. 4(b). The decreasing energy barrier and the higher polarization values reduce the $E_\mathrm{c}$ of the SA pathway by 14 %, from 8 MV/cm for undoped hafnia to 6 MV/cm at 0.2 holes/f.u. Meanwhile, in the SI pathway, the negligible change of the energy barrier combined with the decrease in $\boldsymbol{P_s}$, results in a slight increase (< 2 MV/cm) in the $E_C$. Thus, the SA pathway becomes the preferred pathway for polarization switching for doping concentrations above 0.15 holes/f.u. (corresponding to ~10$^{22}$ h/cm$^3$). Another consequence of activating the SA pathway is an increase in $\boldsymbol{P_s}$ from 50 μC/cm$^2$ to 70 μC/cm$^2$ after doping of 0.15 h/f.u. Moreover, as previous reports have established [4,7], the sign of the polarization in both pathways are opposite to each other, and thus changing the switching pathway would also lead to a reversal in the polarization direction. The consequence of this sign reversal can lead to opposite movements of the domain walls as described by Wu et al.[4], as well as a change in the sign of the piezoelectric coefficients [7].

Previously, it has been reported that holes in the polar *o*-1 phase of hafnia localize onto the three-fold coordinated $O_I$ sub-lattice, and lowers the electrostatic energy of the system compared to the ground-state monoclinic phase [9]. We therefore examined the localization of the holes across the different phases (polar $Pca2_1$, non-polar orthorhombic *Pbcm*, and tetragonal $P4_2/nmc$) by analyzing the atom-projected density of states and correlated them with the changes in the switching barrier. The holes localize differently in the three phases, as shown in Fig. S2 in the Supplemental Material [41]. In the polar *o*-1 phase, the holes accumulate on the three-fold coordinated $O_I$ ions, as quantified in Fig. S11. In contrast, in the *o*-4 phase, the holes preferentially localize on the four-fold coordinated $O_{II}$ ions, while in the *t*-phase the holes are distributed over the oxygen sub-lattices. For the SI pathway, shown in Fig. S12 in green, the holes are distributed across both the sub-lattices in the *t*-phase and then localize back on the $O_I$ sub-lattice as the polarization direction is reversed. In the SA pathway (in orange), however, holes can hop as a polaron between the two oxygen sub-lattices — that is, from $O_I$ to $O_{II}$ and back to $O_I$ without delocalizing over both the sub-lattices. Kenna et al. [57] have predicted that 2D-polarons could form on the $O_I$ ions due to the flat valence bands, but would be difficult to form on the $O_{II}$ ions in the monoclinic phase. However, the polaronic behavior in other polymorphs of $HfO_2$ have not been reported, and the study of polaron formation energies in the *o*-4 and *t*-phases could shed more light on the trends observed here.

## IV. Discussion

In the following, we discuss the validity of the doping levels considered here by comparing them with previous experimental reports and discuss the implications of activating the SA pathway with hole doping. The doping range discussed in this work can be achieved with chemical dopants. Li et al. [25], Xu et al. [5], and Schroeder et al. [53] were able to grow samples with 10% La doping, 12.5%Y doping and 12-16% La doping, respectively. All these samples retained their polarization and exhibited a ferroelectric hysteresis loop.

The hole concentrations at which the SA pathway becomes activated, 0.15–0.2 h/f.u., correspond to volumetric carrier densities of $(4.45–5.93) \times 10^{21}$ h/cm³. This estimate is obtained by converting the number of holes per formula unit into a volume density using the unit-cell volume of the polar phase, equaling ~125 Å$^3$. For a thin film geometry, these values translate to a 2D carrier density on the order of ~$10^{13}$ h/cm² when assuming a film thickness of 1 nm. Carrier concentrations of this magnitude are experimentally accessible in insulating oxides through electrostatic doping. Electrostatic doping refers to approaches that modify electron or hole populations without chemical substitution [58], where the carrier density is governed by electrostatic interactions between a semiconductor and a different material at the interface. Such electrostatic gating methods, can achieve 2D carrier densities up to $10^{14}$ charge carriers/cm$^2$ [58] for thin films, while interfacial charge accumulation in oxide heterojunctions is commonly reported to reach ~$10^{13}$ charge carriers/cm² [59,60].

In the specific case of $(Hf,Zr)O_2$ thin films, Shi et al. [19] reported that interfacial charge transfer from a $La_{0.67}Sr_{0.33}MnO_3$ substrate stabilizes the o-1 phase. Their theoretical analysis supports the experimental observation, showing that stabilization occurs for hole concentrations exceeding 0.15 h/f.u. These values fall directly within the range predicted here to activate the SA pathway, indicating that substrate-induced interfacial charge transfer can realistically provide the carrier densities required to trigger this transformation mechanism.

Charges can also form intrinsically due to charged domain walls between oppositely polarized domains in ferroelectric systems. Charge concentrations >$10^{20}$ h/cm$^3$ can be reached in such samples with thin wall widths and sizeable polarization [20]. Recently, Noor et al. [61] theoretically proposed that charged domain walls can exist in $HfO_2$. Experimentally, ultrathin charged domain walls have also been reported in the isostructural binary oxide, $ZrO_2$ [29], wherein the charge density can reach $10^{23}$ h/cm$^3$. Such large hole doping concentrations can help stabilize the polar phase locally and also alter the switching pathway of the domains.

We also calculate the change in internal energy due to the introduction of the positive charges. The internal energy change provides a measure of the electrostatic cost associated with introducing holes into the lattice and can indicate whether the doping concentrations are accessible. Since, the introduced holes do not localize into a defect site, conventional charge correction schemes are difficult to implement. We therefore use Ewald energy to quantify the change in the electrostatic energy, following Cao et al. [9], and provide an estimate for the change in bulk energy upon introduction of the holes. Further details are provided in Fig. S12. The change in electrostatic energy at 0.15 h/f.u. is 0.6 eV.

Although our discussion focuses on uniform switching—where all unit cells switch polarization simultaneously—it should be noted that ferroelectric switching typically occurs via domain wall motion. Thus, the calculated $E_c$ in our study, which is in the range of 6 – 12 MV/cm is an upper bound. The $E_c$ values reported experimentally lie between 1 – 4 MV/cm [5]. The theoretical overestimation could also be due to a combination of other external factors in experimental samples including strain effects, defects, grain boundaries, temperature and domain characteristics, all of which can lower the coercive field [62]. Nonetheless, the trends with hole doping — i.e., a reduction of the $E_c$ and activation of the SA pathway—are still meaningful. Silva et al. [23] have also theoretically predicted that polycrystalline samples containing interfaces between the monoclinic phase and the *o*-1 phase will have a lower coercive field due to La doping. We anticipate this behavior to extend to single crystalline, hole-doped samples, with commonly observed 180° domains [5,12,25], and to other charged domain configurations [17]. Thus, we hope that the present work motivates mechanistic studies on the effect of hole doping on domain wall switching in experimental studies.

We studied the effect of dopant concentrations on the phase stability of the *o*-4 phase, and understand the effect of chemical dopants on the SA pathway. We substituted Hf with Y and La, both hole dopants widely used in experimental studies on $HfO_2$ [5,53], between concentrations of 6-18 at.%. Both volume and ionic relaxations were allowed. At 6 at.%, the *o*-4 phase was stable but the relative stability between the *o*-4 phase and *o*-1 phase remained unchanged from the undoped case. Above 6 at. %, the *o*-4 phase relaxed into the *m*-phase. The results are summarized in Fig. S6. Thus, the volume expansion created by these dopants influences the stability of the *o*-4 phase and thereby also the energy barrier of the SA pathway. Higher doping concentrations require larger supercells to avoid spurious interactions and the distance between dopants also widely changes the stability of phases in $HfO_2$, as has been reported by Gupta et al [55]. Accordingly, more extensive studies are needed to clarify the influence of chemical dopants on the SA pathway. The combined effects of strain and doping also require further investigation, and our current findings are therefore most applicable to electrostatic doping and charged-domain-wall scenarios.

Our results on the effect of doping on the SI pathway are in good agreement with recent computational reports [55,56]. Gupta et al. [55] computationally investigated the effect of dopants (both *p*-type and *n*-type) and their ionic radii on the $E_c$ for the SI pathway. They found that Y, La, Sr at 6.25% per formula unit increases the $E_c$ by less than 2 MV/cm, in agreement with our results. They also showed that dopants with smaller radii stabilize the *t*-phase and reduce the $E_c$. In contrast, our results show that hole doping produces a negligible change in the relative stability of the *t*-phase with respect to the *o*-1 phase. In their work, the varying atomic radii of the dopants introduces strain in the system, which stabilized the *t*-phase. As discussed above, this strain effect is not seen with the intrinsic charge doping studies performed here. Similarly, Li et al. [56] reported a reduction in the $E_c$ due to both *p*-type and *n*-type chemical doping and identified a previously unreported pathway through the *Pcca* phase. The proposed pathway arises from the SI pathway with the *Pcca* phase formed by freezing the non-polar $M_3$ mode, as discussed in Fig. 1(b) and Fig.

S4 in the Supplemental Material [41] . The *Pcca* phase is however higher in energy than the *t*-phase, as shown in the right panel of Fig. S8(a), and less likely to be a reference phase for switching.

## V. Conclusions

The findings presented in this study highlight the role of hole doping in modulating the polarization switching pathways in $HfO_2$ with an emphasis on the reduction in the coercive field ($E_C$). The reduction of the coercive field arises from the stabilization of the unstable polar mode in the *Pbcm* phase, thereby promoting polarization switching through the SA pathway at doping concentrations higher than 0.15 h/f.u. The activation of the SA pathway can also lead to a change in the polarization direction, which reverses the direction of domain wall movement and the sign of the piezoelectric coefficient. In contrast, the SI pathway remains largely unaffected by hole doping due to the nature of coupling of the various hard phonon modes. The invariance caused by the improper ferroelectric nature of the SI pathway allows for the introduction of charge carriers in $HfO_2$ without a discernible change in its ferroelectric properties. Moreover, as most samples of ferroelectric hafnia are epitaxially grown, the coupling of strain and doping will also effect the switching pathways, as has been predicted for the SI pathway in a recent report [55]. It remains to be investigated whether a similar synergistic coupling of holes and strain could also lower the coercive field for the SA pathway. Beyond chemical doping, holes introduced through electrostatic gating or due to the presence of charged domains walls can lower the coercive field. Overall, our results highlight the need for atomic-scale mechanistic studies on the role of hole dopants on the nucleation and growth of domains in $HfO_2$.

**Data Availability**: The data that support the results of this study are openly available at [63].

**Acknowledgements:** This work was primarily supported the National Science Foundation (NSF) through grant nos. DMR-2145797 (P.O., G.Y.J., R.M.) and DMR-2122070 (G.R., R.M.). This work used computational resources through allocation DMR160007 from the Advanced Cyberinfrastructure Coordination Ecosystem: Services & Support (ACCESS) program, which is supported by NSF grants #2138259, #2138286, #2138307, #2137603, and #2138296.

## VI. References

[1] T. Mikolajick, S. Slesazeck, M. H. Park, and U. Schroeder, Mrs Bulletin **43**, 340 (2018).
[2] T. S. Böscke, J. Müller, D. Bräuhaus, U. Schröder, and U. Böttger, Applied Physics Letters **99** (2011).
[3] D. H. Choe, S. Kim, T. Moon, S. Jo, H. Bae, S. G. Nam, Y. S. Lee, and J. Heo, Materials Today **50**, 8 (2021).
[4] Y. Wu, Y. Zhang, J. Jiang, L. Jiang, M. Tang, Y. Zhou, M. Liao, Q. Yang, and E. Y. Tsymbal, Physical Review Letters **131**, 226802 (2023).
[5] X. Xu, F. T. Huang, Y. Qi, S. Singh, K. M. Rabe, D. Obeysekera, J. Yang, M. W. Chu, and S. W. Cheong, Nat Mater **20**, 826 (2021).
[6] W. Wei *et al.*, Journal of Applied Physics **131**, 154101 (2022).
[7] Y. Qi, S. E. Reyes-Lillo, and K. M. Rabe, arXiv preprint arXiv:2204.06999 (2022).
[8] X. Li *et al.*, 2024), p. arXiv:2408.01830.
[9] T. F. Cao, G. D. Ren, D. F. Shao, E. Y. Tsymbal, and R. Mishra, Physical Review Materials **7**, 044412 (2023).
[10] R. Materlik, C. Künneth, M. Falkowski, T. Mikolajick, and A. Kersch, Journal of Applied Physics **123** (2018).
[11] T. F. Song, H. Tan, A. C. Robert, S. Estandia, J. Gázquez, F. Sánchez, and I. Fina, Applied Materials Today **29**, 101621 (2022).
[12] Y. Yun *et al.*, Nat Mater **21**, 903 (2022).
[13] T. Song, S. Estandía, I. Fina, and F. Sánchez, Applied Materials Today **29**, 101661 (2022).
[14] T. F. Song, R. Bachelet, G. Saint-Girons, R. Solanas, I. Fina, and F. Sánchez, Acs Applied Electronic Materials **2**, 3221 (2020).
[15] J. Jeong, Y. Han, and H. Sohn, Journal of Alloys and Compounds **927**, 166961 (2022).
[16] T. V. Perevalov and D. R. Islamov, Computational Materials Science **233**, 112708 (2024).
[17] Z. S. Xu, X. A. Zhu, G. D. Zhao, D. W. Zhang, and S. F. Yu, Applied Physics Letters **124** (2024).
[18] L. Y. Ma and S. Liu, Phys Rev Lett **130**, 096801 (2023).
[19] S. Shi *et al.*, Phys Rev Lett **133**, 036202 (2024).
[20] P. S. Bednyakov, B. I. Sturman, T. Sluka, A. K. Tagantsev, and P. V. Yudin, Npj Computational Materials **4**, 65 (2018).
[21] A. Crassous, T. Sluka, A. K. Tagantsev, and N. Setter, Nature Nanotechnology **10**, 614 (2015).

[22] Y. Wu, Y. Zhang, J. Jiang, L. Jiang, M. Tang, Y. Zhou, E. Y. Tsymbal, M. Liao, and Q. Yang, arXiv preprint arXiv:2301.06248 (2023).
[23] A. Silva, I. Fina, F. Sánchez, J. P. B. Silva, L. Marques, and V. Lenzi, Materials Today Physics **34**, 101064 (2023).
[24] Y. Cheng *et al.*, Nat Commun **13**, 645 (2022).
[25] X. Li *et al.*, arXiv preprint arXiv:2408.01830 (2024).
[26] K. Chae, A. C. Kummel, and K. Cho, ACS Appl Mater Interfaces **14**, 29007 (2022).
[27] R. He, H. Y. Wu, S. Liu, H. F. Liu, and Z. C. Zhong, Physical Review B **104**, L180102 (2021).
[28] D. R. Islamov and T. V. Perevalov, Microelectronic Engineering **216**, 111041 (2019).
[29] N. Afroze *et al.*, 2025), p. arXiv:2507.18920.
[30] P. Jiang *et al.*, Advanced Electronic Materials **7** (2020).
[31] G. Kresse and J. Furthmuller, Physical review B **54**, 11169 (1996).
[32] J. P. Perdew, K. Burke, and M. Ernzerhof, Physical Review Letters **77**, 3865 (1996).
[33] P. E. Blöchl, Physical Review B **50**, 17953 (1994).
[34] S. P. Ong *et al.*, Computational Materials Science **68**, 314 (2013).
[35] K. Momma and F. Izumi, J Appl Crystallogr **41**, 653 (2008).
[36] M. Leslie and N. J. Gillan, Journal of Physics C: Solid State Physics **18**, 973 (1985).
[37] M. I. Aroyo, J. M. Perez-Mato, D. Orobengoa, E. Tasci, G. de la Flor, and A. Kirov, Bulgarian Chemical Communications **43**, 183 (2011).
[38] H. Stokes, D. Hatch, and B. Campbell, Google Scholar There is no corresponding record for this reference (2021).
[39] S. Li and T. Birol, Phys Rev Lett **127**, 087601 (2021).
[40] S. E. Reyes-Lillo, K. F. Garrity, and K. M. Rabe, Physical Review B **90**, 140103 (2014).
[41] See Supplemental Material [Insert URL] for complementary calculations. Table S1: The values of the fitting coefficients of SI pathway for different hole-doping levels. Fig. S5: The difference in the displacements of the three-fold coordinated oxygen ions. Fig. S2: Density of states projected the Hf, $O_I$ and $O_{II}$ sublattices. Fig. S3: The change in energy barriers for $ZrO_2$. Fig. S4: Symmetry mode relationships between the high symmetry non-polar phases, with the lower symmetry phases. Fig. S5: Relative energies of the possible $HfO_2$ polymorphs as a function of hole concentration. Fig. S6: Energy difference between the *Pbcm* and $Pca2_1$ phase with increasing concentration of substitutional Y and La doping at the Hf-site. Fig. S7: The value of the coefficients in the Landau equation for the SA pathway. Fig. S8: The change in energy for various modes and their couplings in the SI pathway. Fig. S9: Solid State - Nudged Elastic Band (SS-NEB) calculations for both pathways. Fig. S10: The magnitude of the polar distortion along the c-axis for both pathways. Fig. S11: The amount of holes localized on the $O_I$ sublattice as a function of total hole concentration. Fig. S12: The change in internal bulk energy as a function of hole concentration. Fig. S13: The localization of holes across different oxygen sublattices for SI and SA pathways.
[42] F. Delodovici, P. Barone, and S. Picozzi, Physical Review Materials **5**, 064405 (2021).
[43] H. Aramberri and J. Iñiguez, Communications Materials **4** (2023).
[44] H. J. Lee, M. Lee, K. Lee, J. Jo, H. Yang, Y. Kim, S. C. Chae, U. Waghmare, and J. H. Lee, Science **369**, 1343 (2020).
[45] S. T. Hartman, A. S. Thind, and R. Mishra, Chemistry of Materials **32**, 9542 (2020).
[46] N. A. Benedek and C. J. Fennie, Phys Rev Lett **106**, 107204 (2011).
[47] J. H. Lee *et al.*, Nature **466**, 954 (2010).
[48] J. Nordlander, M. Campanini, M. D. Rossell, R. Erni, Q. N. Meier, A. Cano, N. A. Spaldin, M. Fiebig, and M. Trassin, Nat Commun **10**, 5591 (2019).
[49] Y. Yun, P. Buragohain, A. S. Thind, Y. Yin, X. Li, X. Jiang, R. Mishra, A. Gruverman, and X. Xu, Physical Review Applied **18**, 034071 (2022).
[50] J. Junquera and P. Ghosez, Nature **422**, 506 (2003).

[51] M. Lederer, R. Olivo, D. Lehninger, S. Abdulazhanov, T. Kämpfe, S. Kirbach, C. Mart, K. Seidel, and L. M. Eng, physica status solidi (RRL)–Rapid Research Letters **15**, 2100086 (2021).
[52] Y. Tan, Y. Wu, T. Duan, R. Xiong, J. Jiang, M. Liao, and Q. Yang, Computational Materials Science **253**, 113873 (2025).
[53] U. Schroeder *et al.*, Inorg Chem **57**, 2752 (2018).
[54] S. Starschich, S. Menzel, and U. Böttger, Journal of Applied Physics **121** (2017).
[55] D. Gupta, P. Kumar, and J. H. Lee, Physical Review Applied **22** (2024).
[56] G. Li, S. Yan, Y. Liu, W. Zhang, Y. Xiao, Q. Yang, M. Tang, J. Li, and Z. Long, npj Computational Materials **11**, 1 (2025).
[57] K. P. McKenna, M. J. Wolf, A. L. Shluger, S. Lany, and A. Zunger, Phys Rev Lett **108**, 116403 (2012).
[58] A. Goldman, Annual Review of Materials Research **44**, 45 (2014).
[59] H. Wang *et al.*, ACS nano **14**, 12697 (2020).
[60] H. Hahn, B. Pécz, A. Kovács, M. Heuken, H. Kalisch, and A. Vescan, Journal of applied physics **117** (2015).
[61] M. Noor, M. Bergschneider, J. Kim, N. Afroze, A. I. Khan, S. C. Chang, U. E. Avci, A. C. Kummel, and K. Cho, ACS Appl Mater Interfaces **16**, 62282 (2024).
[62] A. N. Morozovska and E. A. Eliseev, Physica B: Condensed Matter **355**, 236 (2005).
[63] Omprakash, Pravan. Hole-doping Reduces the Coercive Field in Ferroelectric Hafnia. Zenodo, August 3, 2025. https://doi.org/10.5281/zenodo.16730004.

# Supplemental Material

## Hole-doping reduces the coercive field in ferroelectric hafnia

Pravan Omprakash[1,*], Gwan Yeong Jung[2], Guodong Ren[1], Rohan Mishra[2,1,†]

[1]*Department of Mechanical Engineering & Material Science, Washington University in St. Louis, St. Louis, MO 63130, USA*

[2]*Institute of Materials Science & Engineering, Washington University in St. Louis, St. Louis, MO 63130, USA*

[*]Corresponding author: o.pravan@wustl.edu

[†]Corresponding author: rmishra@wustl.edu

**Table S1.** The values of tl · different hole-doping levels.

| h/f.u. | $a_{200}$ | $a_{400}$ | $a_{600}$ | $a_{020}$ | $a_{040}$ | $a_{060}$ | $a_{002}$ | $a_{004}$ | $a_{006}$ | $a_{220}$ | $a_{240}$ | $a_{420}$ | $a_{022}$ | $a_{024}$ | $a_{042}$ | $a_{202}$ | $a_{204}$ | $a_{402}$ | $a_{111}$ | $a_{222}$ | $a_{113}$ | $a_{131}$ | $a_{311}$ |
|---|---|---|---|---|---|---|---|---|---|---|---|---|---|---|---|---|---|---|---|---|---|---|---|
| **0** | 27 | 45 | 0 | 54 | 16 | 6 | 194 | -16 | 3 | 326 | -215 | 205 | 233 | 298 | -253 | 148 | -52 | 78 | -1033 | -140 | -53 | 190 | -144 |
| **0.04** | 45 | 31 | 2 | 90 | 0 | 8 | 161 | 14 | -3 | 351 | -262 | 242 | 267 | -68 | 107 | 207 | 116 | -99 | -1097 | -81 | -114 | 93 | -90 |
| **0.08** | 68 | 13 | 6 | 129 | -18 | 11 | 145 | 28 | -6 | 358 | -145 | 116 | 280 | 64 | -30 | 244 | 9 | 0 | -1181 | -73 | -25 | -12 | -56 |
| **0.12** | 95 | -4 | 9 | 177 | -26 | 11 | 140 | 43 | -9 | 357 | -382 | 352 | 303 | -68 | 103 | 297 | -123 | 129 | -1297 | -69 | 650 | -752 | -5 |
| **0.16** | 117 | -24 | 13 | 210 | -54 | 15 | 139 | 36 | -8 | 373 | -738 | 689 | 315 | 9 | 9 | 313 | 1 | -13 | -1469 | -113 | -47 | 992 | -849 |
| **0.2** | 152 | -53 | 19 | 247 | -73 | 17 | 147 | 27 | -5 | 367 | -772 | 715 | 316 | -46 | 56 | 335 | -380 | 356 | -1406 | -14 | -1188 | 396 | 743 |

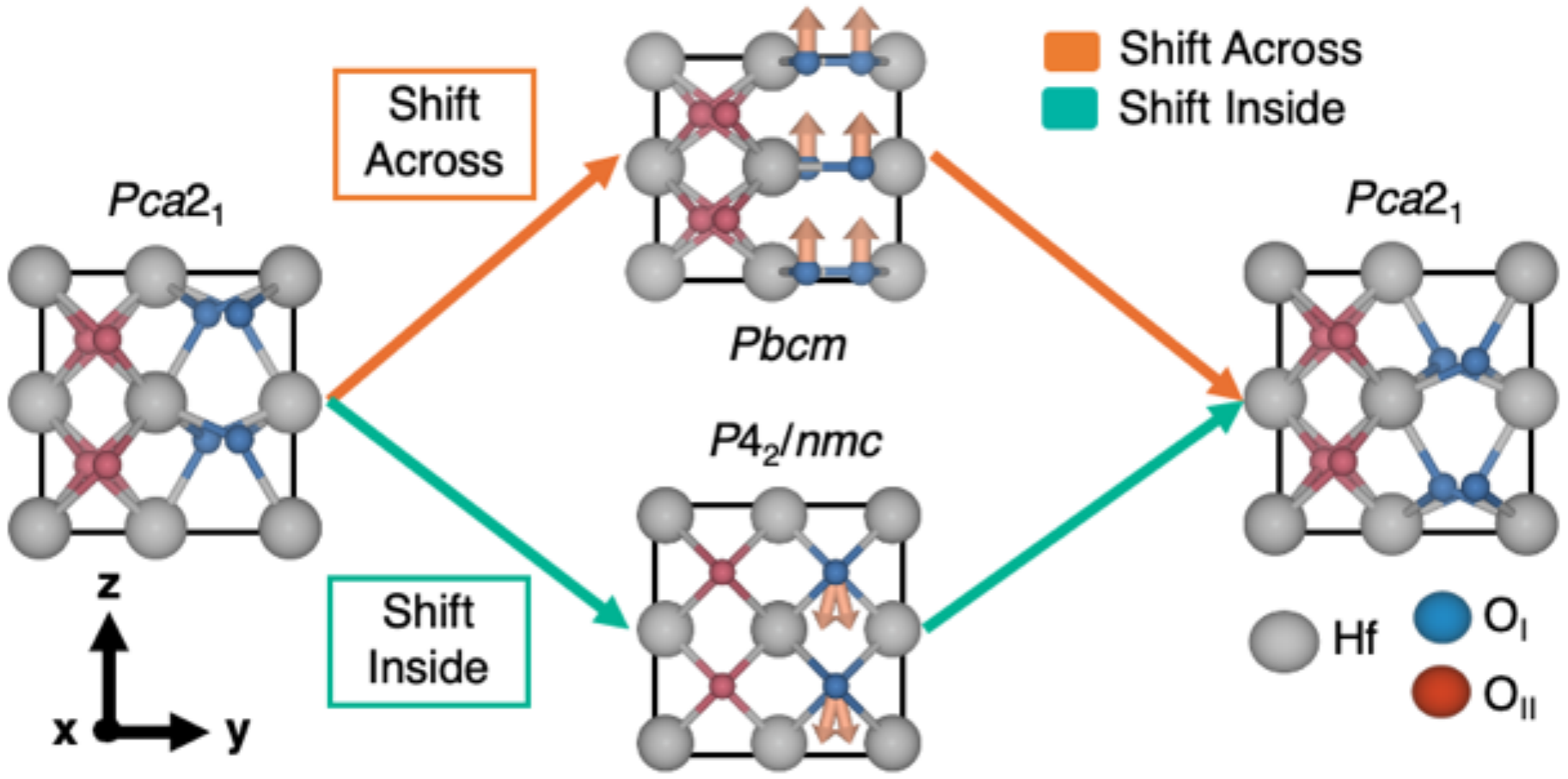

**Fig. S6.** The difference in the displacements of the three-fold coordinated oxygen ions ($O_I$ in blue) between the Shift Across (SA) pathway in orange and the Shift Inside (SI) pathway in green. The $O_I$ ions move along opposite directions leading to differences in polarization.

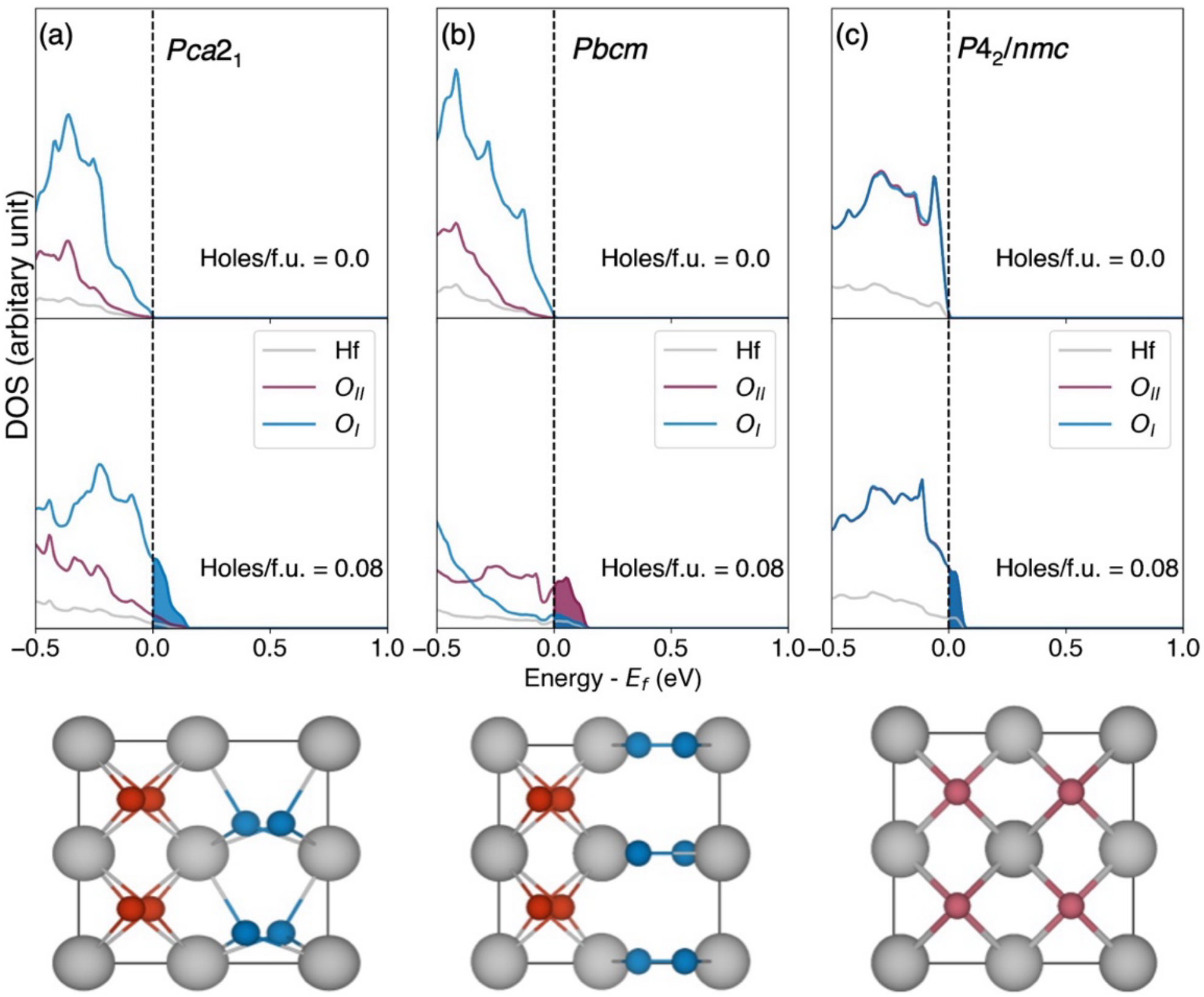

**Fig. S2.** Density of states projected onto the Hf (grey), $O_I$ (blue) and $O_{II}$ (red) sublattices in the (a) $Pca2_1$ (*oI*-phase), (b) *Pbcm* (*oIV*-phase), and in (c) $P4_2/nmc$ (*t*-phase). The shaded states above the Fermi energy, $E_F$, denoted by the vertical black dashed lines, indicate doped holes on the different sublattices.

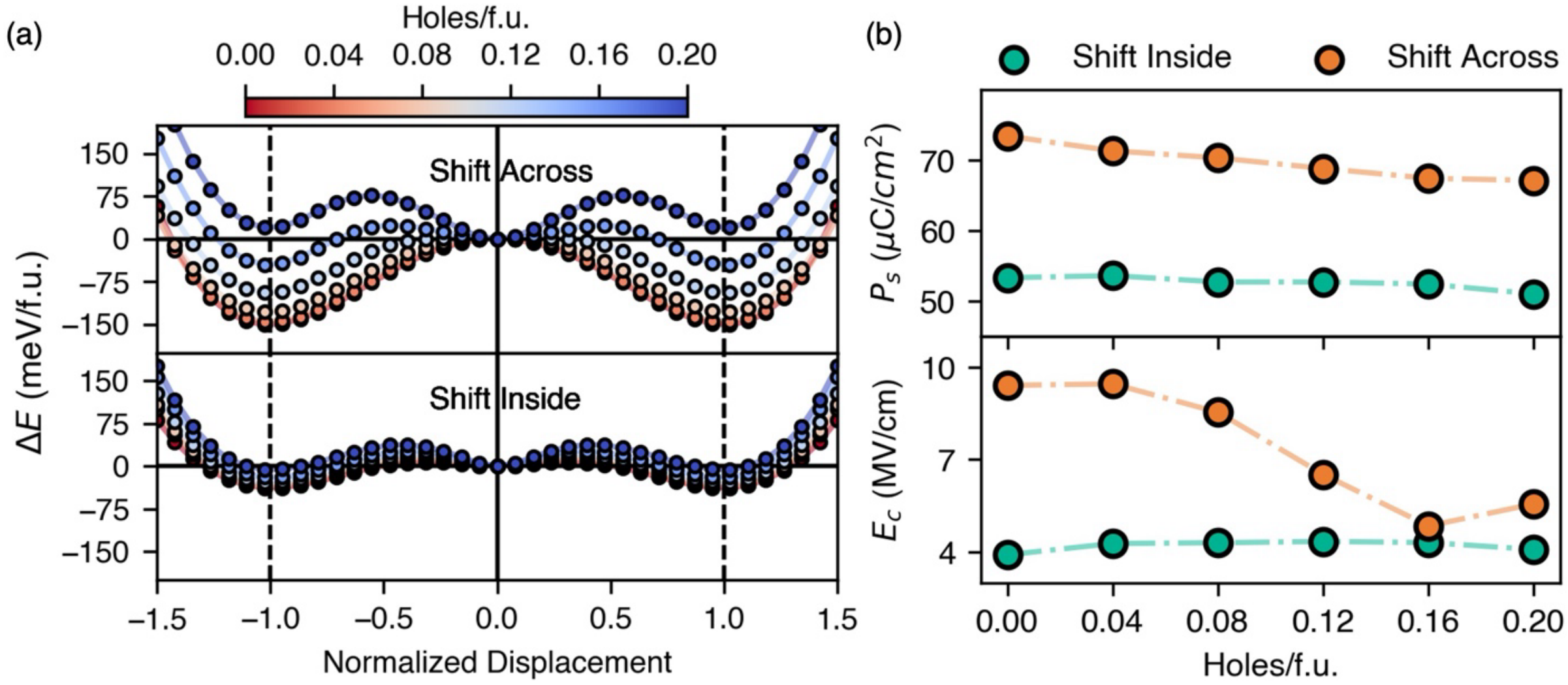

**Fig. S3.** The change in energy barriers for (a) Shift Across (SA) and (b) Shift Inside (SI) pathways with the corresponding (c) spontaneous polarization, *P* (upper panel) and coercive field, $E_C$ (lower panel) as a function of hole doping for $ZrO_2$. The coercive field for the SA pathway decreases just as in $HfO_2$, however the polar phase becomes metastable by 0.16 h/f.u., thus destabilizing ferroelectricity in $ZrO_2$.

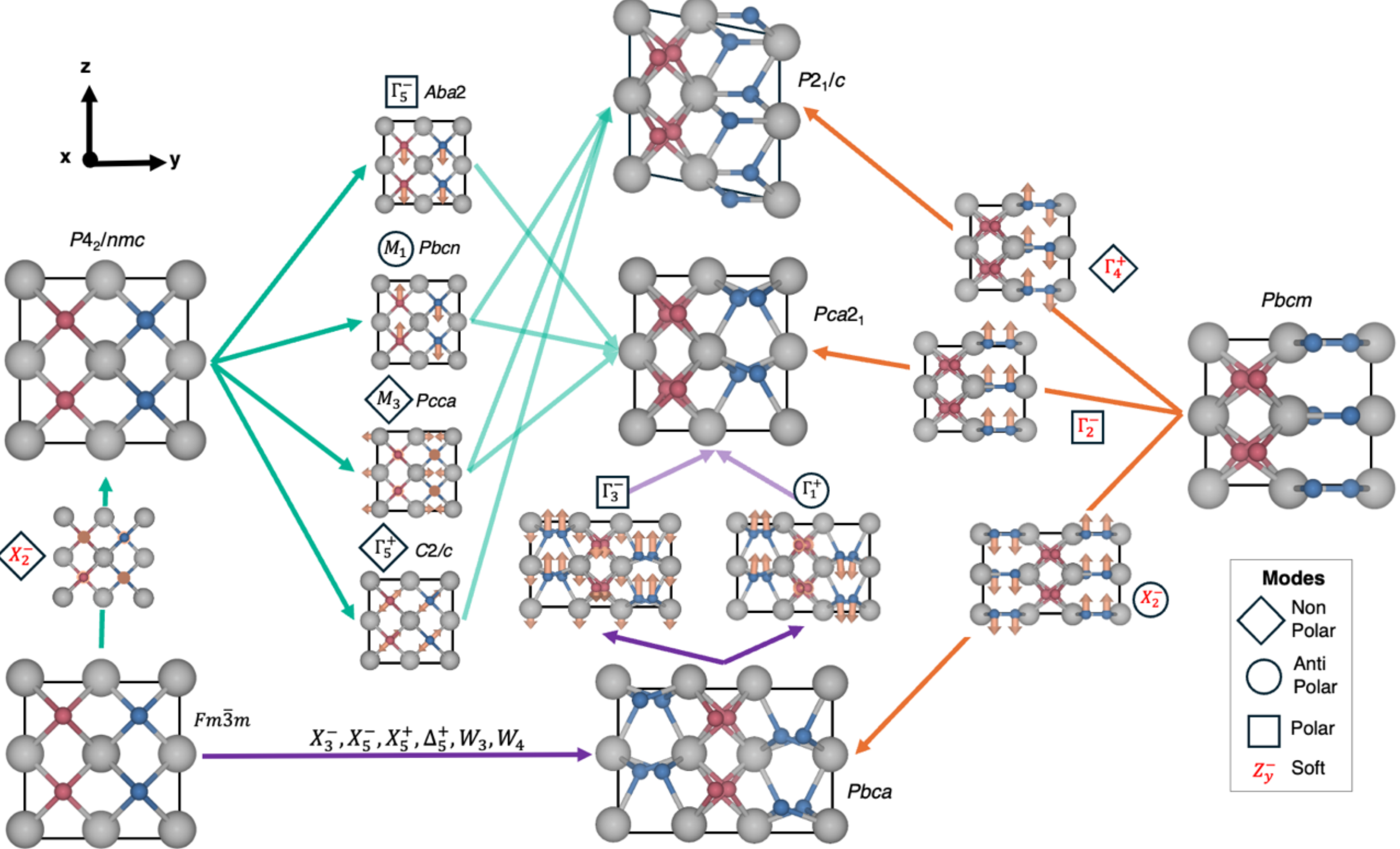

**Fig. S4.** Symmetry mode relationships between the high symmetry non-polar phases (shown on both the right and the left edges), with the lower symmetry phases including the ground state monoclinic $P2_1/c$ phase, the polar $Pca2_1$ phase and the antiferroelectric *Pbca* phase. Transitions from the *Pbcm* (*o*IV-phase) are shown with orange arrows, and consist of three soft modes ($\Gamma_2^-, \Gamma_5^-, X_2^-$). Transitions from the $P4_2/nmc$ (t-phase) are shown with green arrows, and consist of four hard modes, three of which ($\Gamma_5^-$, $M_1$, $M_3$) couple to stabilize the polar $Pca2_1$ phase, and the other three ($\Gamma_5^+$, $M_1$, $M_3$) stabilize the monoclinic $P2_1/c$ phase (*m*-phase). The *Pbca* (*o*II-phase) phase (shown in purple arrows) is symmetry related to only the polar phase through two hard modes ($\Gamma_1^+, \Gamma_3^-$). Additionally, the symmetry relations to the cubic $Fm\bar{3}m$ (*c*-phase) are also shown.

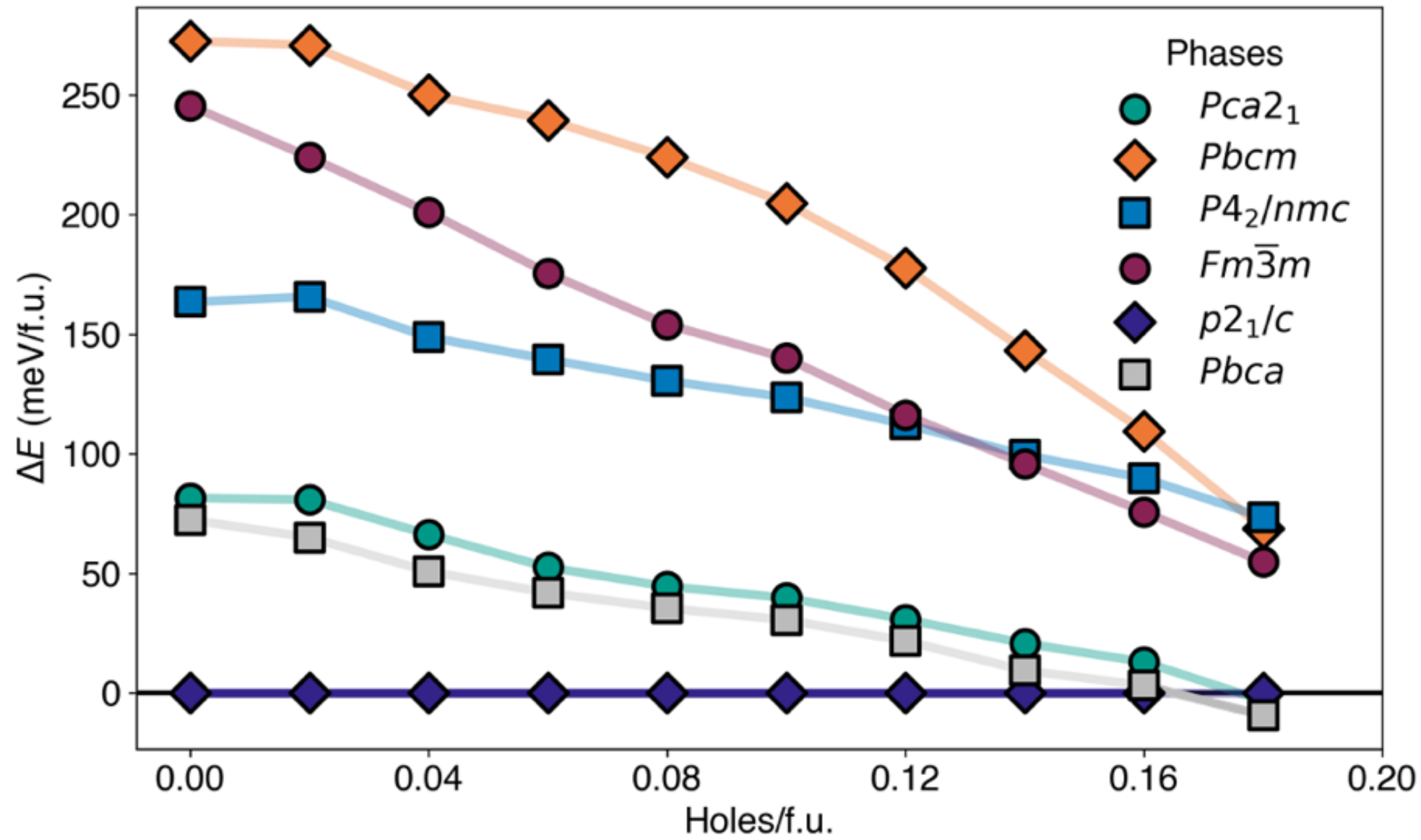

**Fig. S5.** Relative energies ($\Delta E$) of the possible $HfO_2$ polymorphs as a function of hole concentration (holes/f.u.). Both volume and atomic positions are allowed to relax. The energies are referenced to the monoclinic $P2_1/c$ phase (purple) at each doping level. The various polymorphs are shown in Fig. S4.

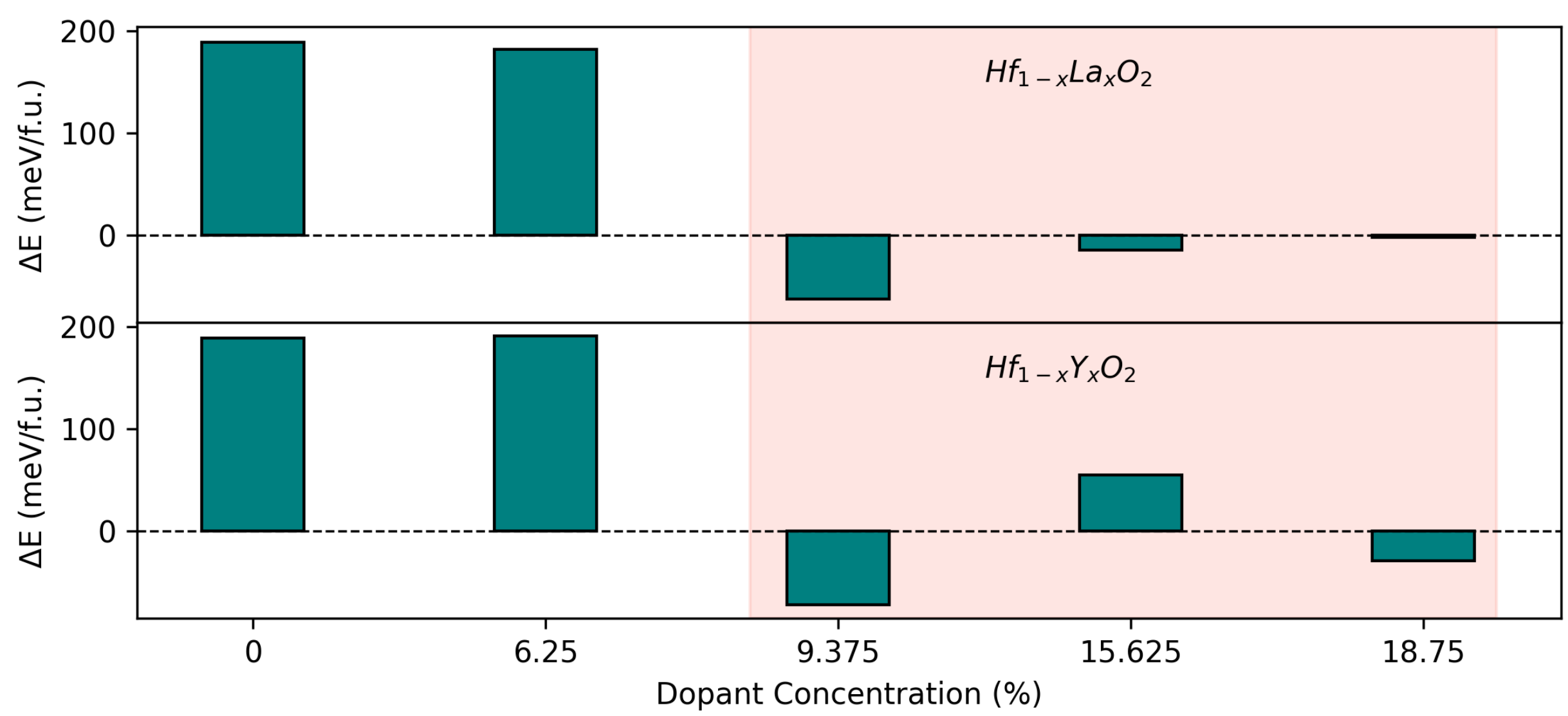

**Fig. S6.** Energy difference between the *Pbcm* and $Pca2_1$ phase with increasing concentration of substitutional Y and La doping at the Hf-site. All calculations were performed on a 2 × 2 × 2 supercell allowing full ionic and volume relaxation. At a dopant concentration of 6.25 at.% there is no change in the relative energy. At concentrations higher than 9%, the *o*-4 phase relaxes into the monoclinic $P2_1/c$ phase (highlighted in pink), highlighting the effect of volume expansion due to the larger dopant sizes of La (1.032 Å) and Y (0.9 Å) compared to Hf (0.76 Å).

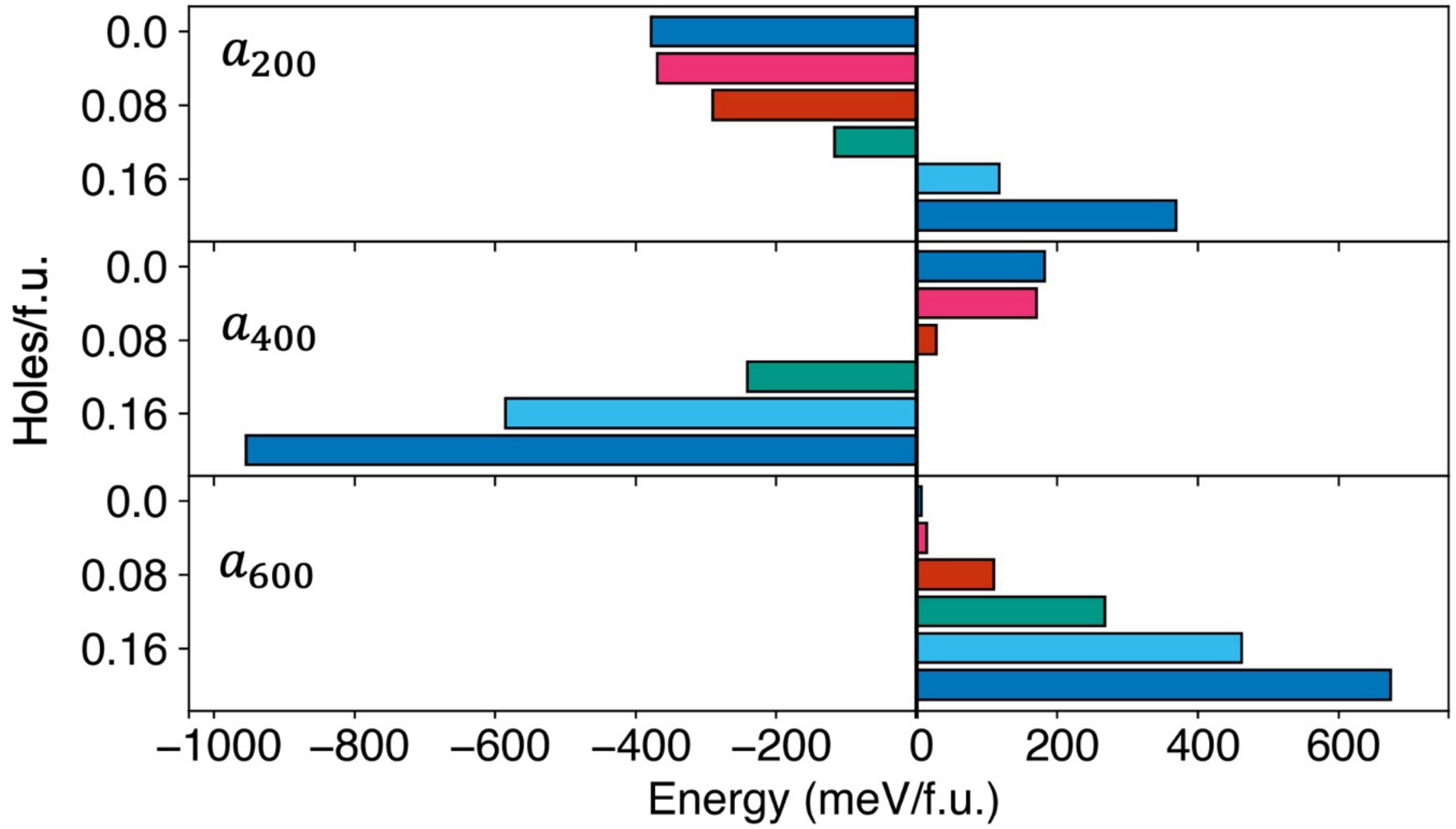

**Fig. S7.** The value of the coefficients in the Landau equation for the SA pathway shown in Eq. (3) in the main text.

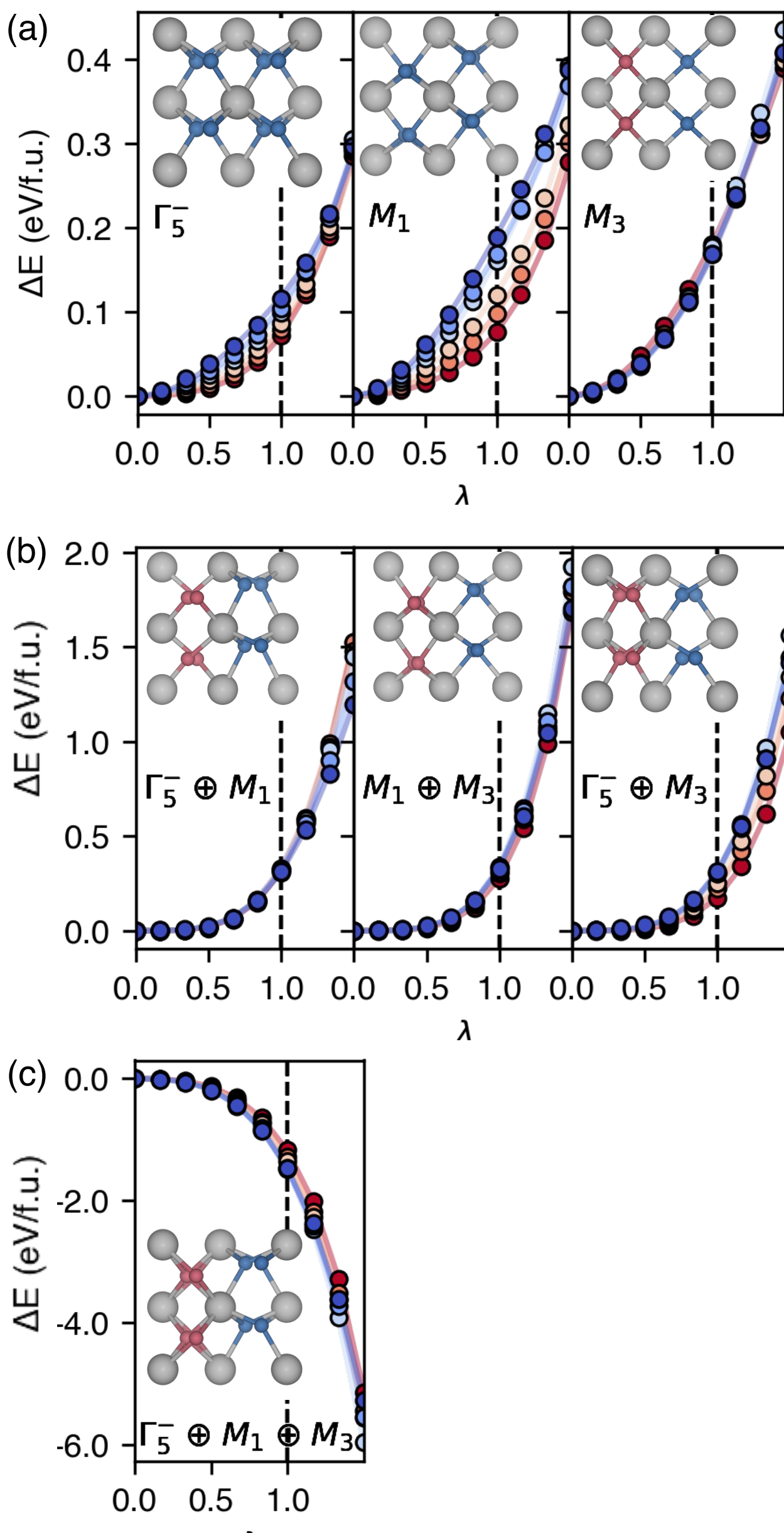

**Fig. S8.** The change in energy for various modes and their couplings in the SI pathway. (a) The three hard modes destabilize the *t*-phase. (b) The three doublet couplings also do not stabilize the phase and (c) the trinomial contribution has a large negative magnitude and stabilizes the polar *o*I-phase.

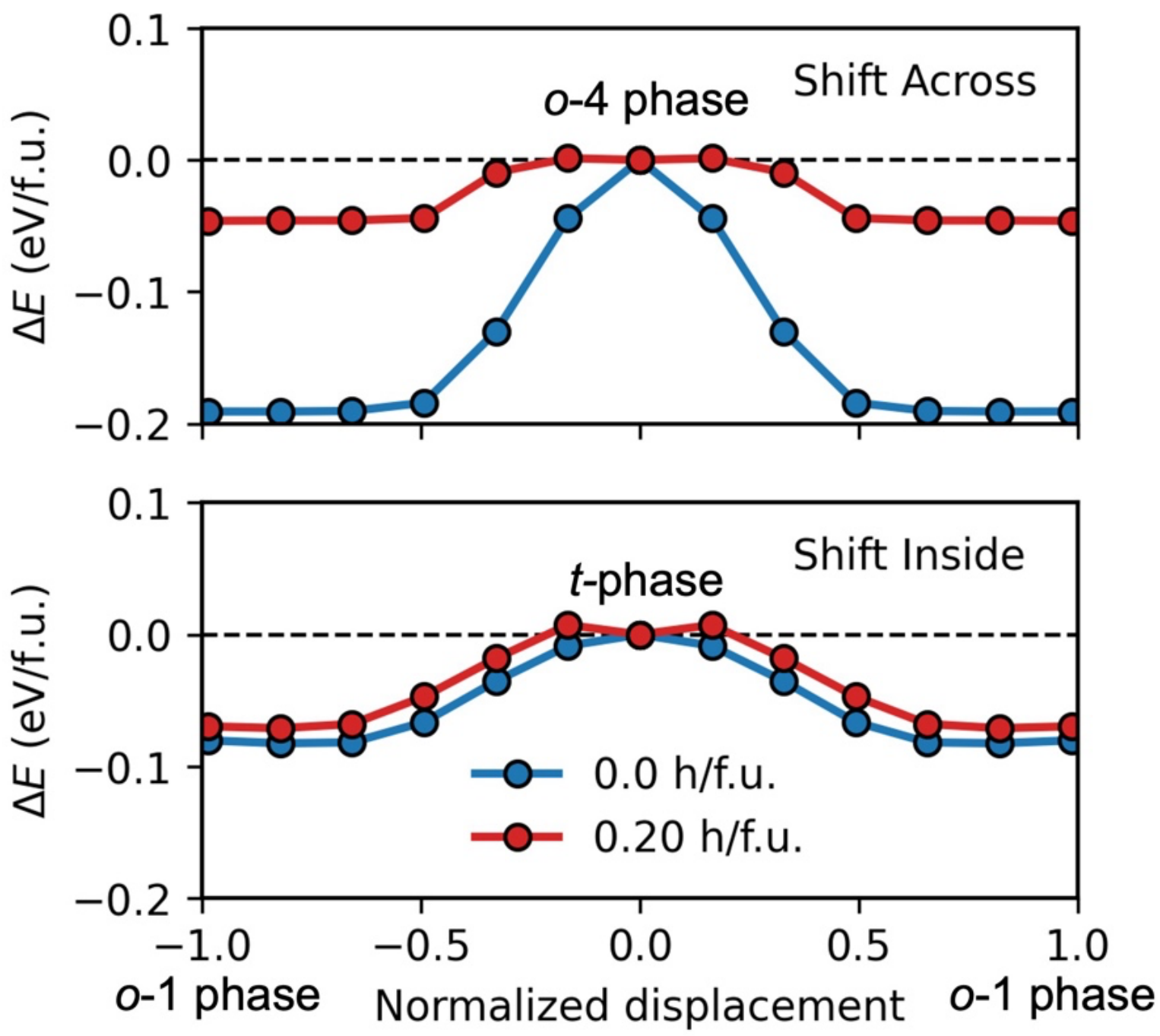

**Fig. S9.** Solid State - Nudged Elastic Band (SS-NEB) calculations for both the Shift Across (top panel) and Shift Inside (bottom panel) pathways at 0.0 and 0.2 h/f.u.. The negligible change in energy barriers for the Shift Inside pathway (lower panel) is consistent with the barriers obtained from symmetry mode analysis. The lowering of the energy barrier in the Shift Across pathway (upper panel) is also consistent between the NEB and the symmetry-mode analysis calculations. The energy barrier is lower for NEB calculations at 0.2 h/f.u than the barrier obtained from symmetry-mode analysis.

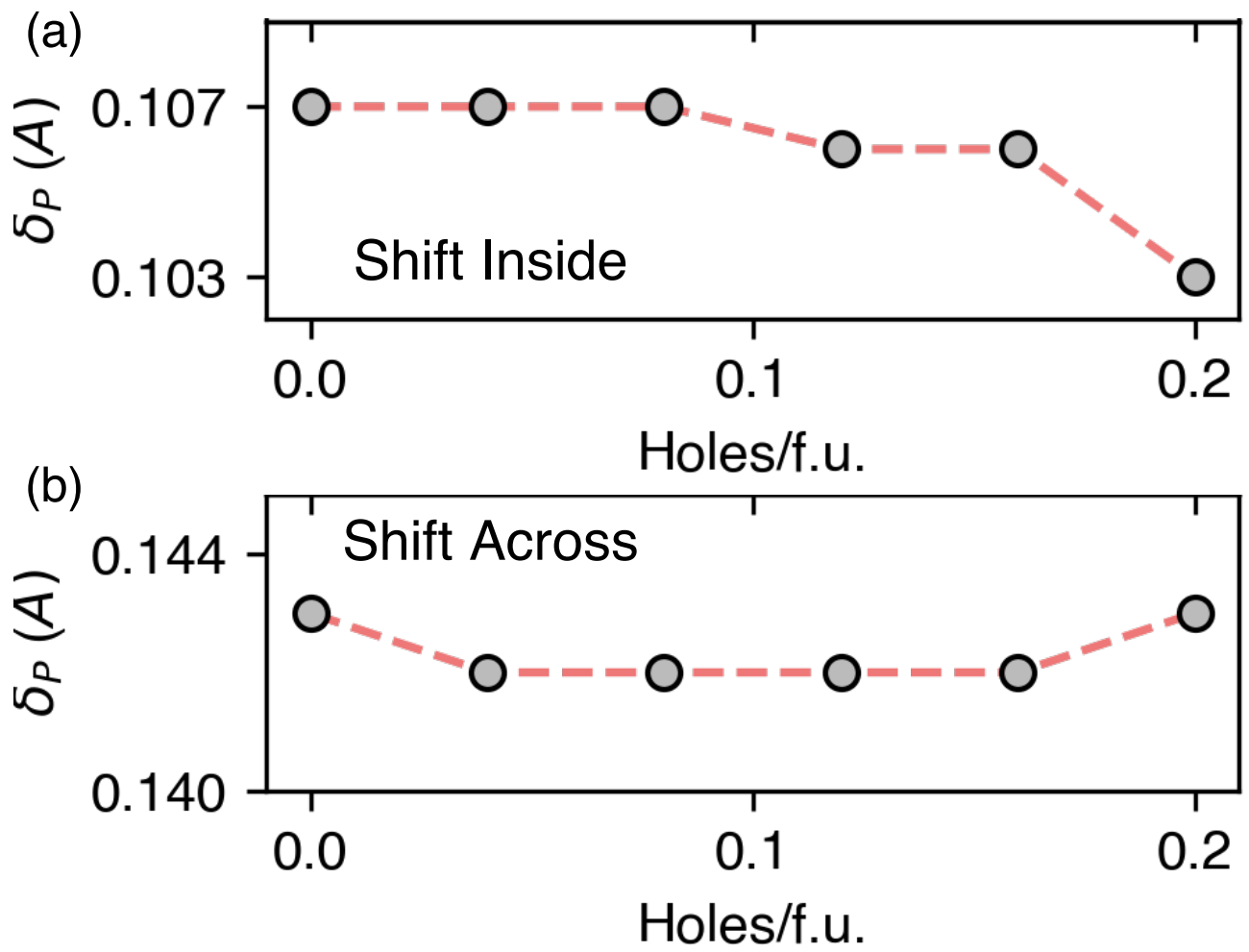

**Fig. S10.** The magnitude of the polar distortion along the c-axis for the (a) SI pathway and (b) SA pathway. These polar distortions were used to spontaneous polarization, $P_s$, using Eq. (1) in the main text.

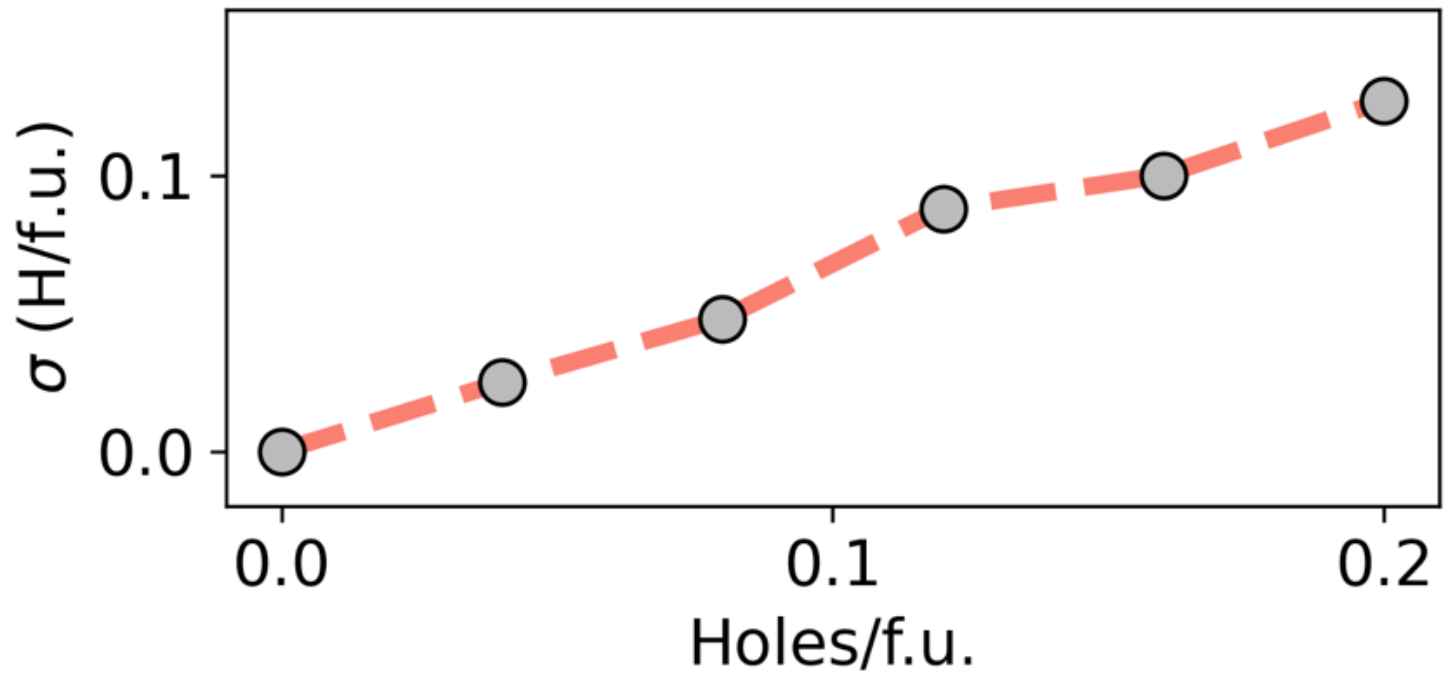

**Fig. S11.** The amount of holes localized on the $O_I$ sublattice as a function of total hole concentration.

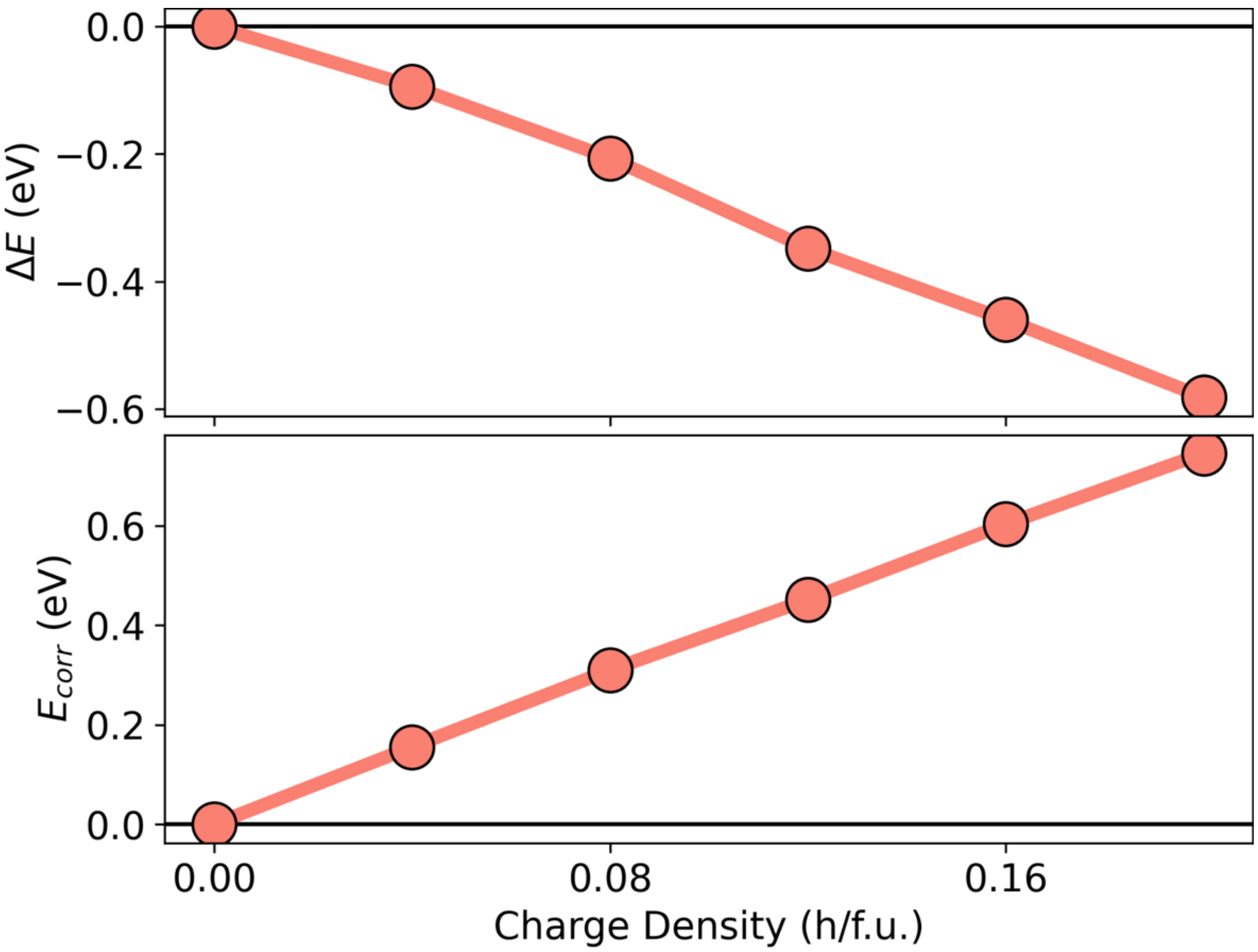

**Fig. S12.** The change in internal bulk energy as a function of hole concentration (top panel) and the energy correction term (bottom panel) introduced to compensate for the applied uniform background charge. $E_{corr}(q) = (E_{ewald}(q) - E_{ewald}(0))/\epsilon$ , where $\epsilon = 23.17$ is the dielectric constant of the *o*-2 phase, $q$ is the charge introduced into the cell, and $E_{ewald}$ is the electrostatic energy between periodic array of charges. We choose the Ewald energy because the added charges are present on all the three-fold coordinated oxygen atoms rather than localizing into a concentrated point defect. The corresponding change in bulk energy is then given by, $\Delta E = E(q) - E_{bulk} + E_{corr}(q)$, where $E_{bulk}$ is the total energy of the undoped phase, and $E(q)$ is the total energy of the charged state. Since the introduced charge forms 2D polarons, and is therefore not a point charge defect, the calculated energies imply an order-of-magnitude estimate for the change in energy when holes are introduced into the system.

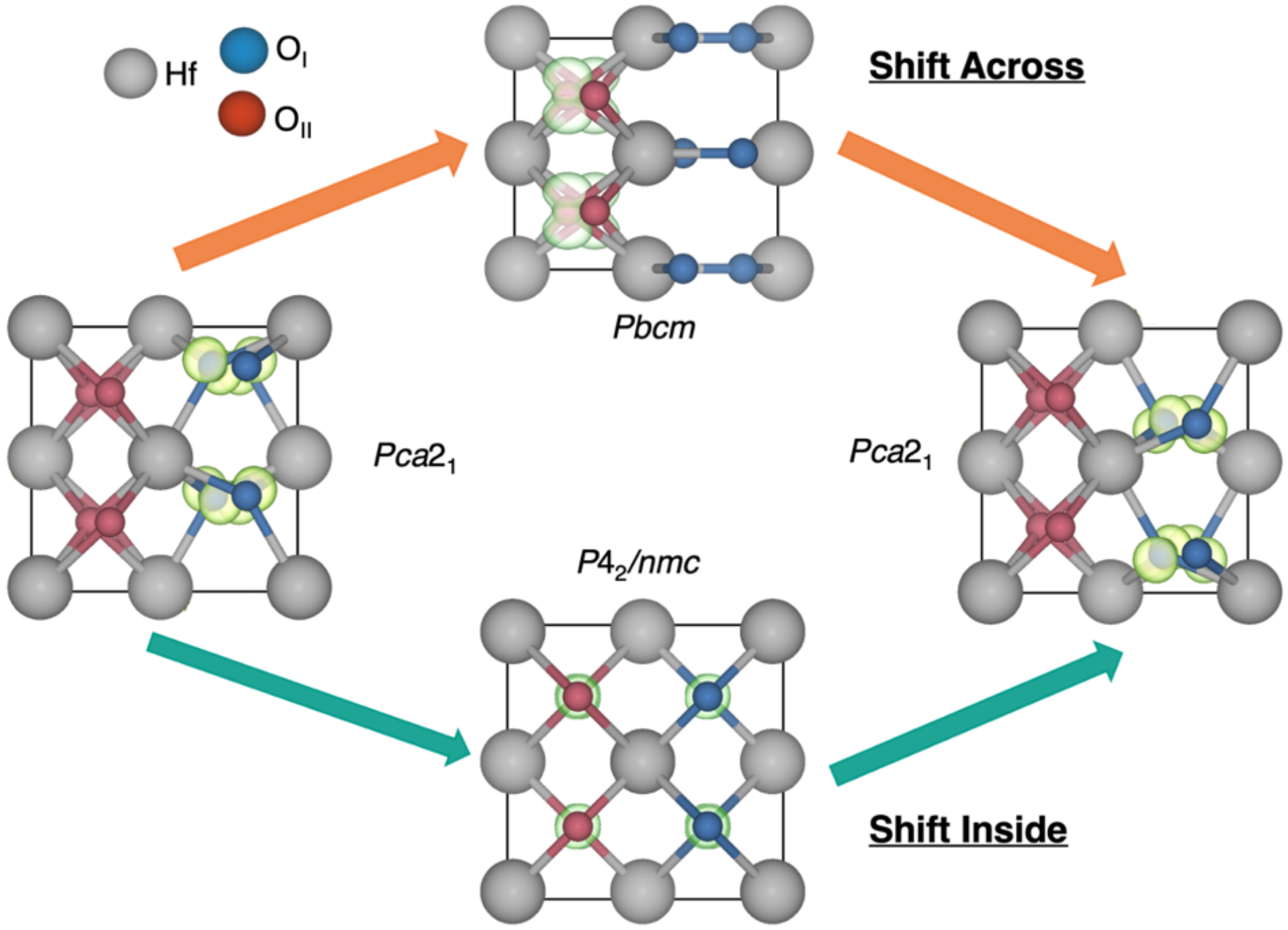

**Fig. S13.** The localization of holes across different oxygen sublattices for SI and SA pathways.